\documentclass{aa}
\usepackage{endnotes}
\let\footnote=\endnote
\usepackage{graphicx}
\usepackage{natbib}
\usepackage{txfonts}
\usepackage{ifpdf}
\pdfoutput=0
\ifpdf
\else

\begin{document}

\title{Wind mapping in Venus' upper mesosphere with the IRAM-Plateau de Bure interferometer}

\author{A. Moullet
          \inst{1,2}
          \and
          E. Lellouch
        \inst{3}
           \and
        R. Moreno
        \inst{3}  
       \and
        M. Gurwell
         \inst{2}
         \and
        H. Sagawa
         \inst{4} }
\institute{National Radio Astronomy Observatory, Charlottesville VA-22902, U.S.A
\and 
 Harvard-Smithsonian Center for Astrophysics, Cambridge MA-02138, U.S.A
      \and
LESIA-Observatoire de Paris, 5 place J. Janssen, 92195 Meudon CEDEX, France
\and
National Institute of Information and Communications Technology, 4-2-1 Nukui-kita, Koganei, Tokyo, Japan}
\date{November 13th, 2011}
\abstract{}{The dynamics of the upper mesosphere of Venus ($\sim$85-115~km) have been characterized as a combination of a retrograde superrotating zonal wind (RSZ) with a subsolar-to-antisolar flow  (SSAS). Numerous mm-wave single-dish observations have been obtained and could directly measure mesospheric line-of-sight winds by mapping Doppler-shifts on CO rotational lines, but their limited spatial resolution makes their interpretation difficult. By using interferometric facilities, one can obtain better resolution on  Doppler-shifts maps, allowing in particular to put firmer constraints on the respective contributions of the SSAS and RSZ circulations to the global mesospheric wind field. }{We report on interferometric observations of the CO(1-0) line obtained with the IRAM-Plateau de Bure interferometer in November 2007 and June 2009, that could map the upper mesosphere dynamics on the morning hemisphere with a very good spatial resolution (3.5-5.5"). }{ All the obtained measurements show, with a remarkably good temporal stability, that the wind globally flows in the (sky) East-West direction, corresponding in the observed geometry either to an unexpected prograde zonal wind or a SSAS flow. A very localized inversion of the wind direction, that could correspond to a RSZ wind, is also repeatedly  detected in the night hemisphere. The presence of significant meridional winds is not evidenced. Using models with different combinations of zonal and SSAS winds, we find that the data is best reproduced by a dominant SSAS flow with a maximal velocity at the terminator of $\sim$200~m/s, displaying large diurnal and latitudinal asymmetries, combined with an equatorial RSZ wind of 70-100~m/s, overall indicating a wind-field structure consistent with but much more complex than the usual representation of the mesospheric dynamics.}{}
\authorrunning{A. Moullet et al.}
\titlerunning{Venus' mesospheric wind mapping with IRAM-PdBI}

\keywords{Planets and satellites: atmospheres}

\maketitle

\section{Introduction}
The bulk of Venus' atmosphere is usually described as the superposition of three different regions, that present significantly different thermal structures and dynamic regimes. The troposphere is the deepest region and extends up to the top of the cloud deck ($\sim$65 km altitude). This region displays a global strong  retrograde superrotating zonal wind (RSZ), whose centrifugal force is believed to balance the latitudinal temperature gradient, thus maintaining a planet-wide cyclostrophic equilibrium. Pioneer and Venera probes revealed a significant increase of the wind velocity with altitude in the troposphere, reaching equatorial velocities V$_{eq}$ up to $\sim$100~m/s in the cloud deck \citep{gierasch1997},  that were confirmed by tracking of ultraviolet cloud features performed by Mariner, Pioneer and Galileo spacecrafts \citep[see review in][]{limaye2007} and, later, the VMC monitoring camera and VIRTIS imaging spectrometer on board Venus Express \citep{markiewicz2007,sanchez2008}.\\
Above 115~km lies the thermosphere, where the temperature distribution is mainly controlled by the solar EUV heating. As a consequence, the very steep temperature gradient between the night and day hemispheres \citep[up to 150~K, see][]{kasprzak1997} should drive a strong sub-solar to anti-solar (SSAS) wind. Models of the thermospheric SSAS wind \citep{bougher1986} have predicted that the flow should be symmetric around the sub-solar/anti-solar line, and that its velocity should increase almost linearly with the solar incidence angle, to reach a maximum velocity V$_{ter}$ near the terminator of up to 230~m/s above 150~km. The presence of a SSAS wind in the thermosphere is supported by the spatial distribution of light species and NO airglows, which are concentrated at post-midnight local times \citep{bougher1997,gerard2008}, also implying the coexistence of an highly temporally variable RSZ wind. The main mechanism invoked to maintain the RSZ wind in the thermosphere is the breaking of gravity waves generated in the cloud deck \citep{bougher1997}. \\
\\
In between these layers lies the mesosphere (65-115~km). Dynamically, this region is viewed as a transition zone, where both SSAS and RSZ winds coexist, as confirmed by numerous observations. At the lower boundary of the mesosphere, corresponding to the top of the cloud deck ($\sim$68~km), the winds' line-of-sight projection can directly be measured from the Doppler-shifts of the reflected solar Fraunhofer lines on the day-side. Such observations by \citet{widemann2007,widemann2008} revealed a strong and dominant RSZ wind with V$_{eq}$ $\sim$100~m/s and possibly meridional winds (of the order of 10~m/s), as well as large temporal variations, although the latter are not seen in cloud tracking observations sounding just a few kilometers below \citep{moissl2009}.\\
Between 85 and 110~km (upper mesosphere), Doppler-shifts measurements can be obtained from mm-wave CO rotational lines or infrared (IR) CO$_2$ non-LTE emission lines observed by heterodyne spectroscopy. Most results have been interpreted as a combination of SSAS and RSZ winds \citep[see e.g.][]{lellouch1994,clancy2008,lellouch2008,sornig2008}, with variable relative contributions : V$_{ter}$ is usually inferred to be of the order of 100~m/s, while V$_{eq}$ varies between 0 and 200~m/s, and may not be sufficient to maintain cyclostrophic equilibrium \citep{piccialli2008,piccialli2010}. Poleward meridional winds may also have been marginally detected \citep{lellouch1994,lellouch2008}. In addition to large temporal variations over timescales of as little as a few hours  \citep{clancy2011}, modifications of the SSAS and RSZ wind-fields with latitude \citep{sornig2008,lellouch2008}, as well as variations of the wind velocity with altitude \citep{lellouch2008,clancy2011} have also been observed. Prograde zonal winds were also recently suggested based on some day-side measurements on the CO(2-1) line by \citet{lellouch2008}. The possible driving mechanism for a prograde zonal wind is unknown. Furthermore, based on the analysis of a large number of night-side observations, \citet{clancy2011} suggested the existence of large-scale circulations that cannot be explained by a combination of RSZ and SSAS winds. \\
The interpretation of Doppler-shifts in terms of wind field is complex due to the difficulty to disentangle the RSZ and SSAS winds depending on the observing geometry and, for mm-wave observations, the lack of spatial resolution available on single-dish facilities. In the 95-115~km altitude region, an indirect method to characterize winds consists in mapping O$_2$, NO and OH airglows \citep[see e.g.][]{gerard2008}. As in the thermosphere, their post-midnight concentration indicates the dominance of the SSAS wind with respect to the RSZ wind. Recently, using the VIRTIS imaging spectrometer on board the European Venus Express spacecraft, \citet{hueso2008} could track O$_2$ airglows confirming the dominant SSAS wind, but also displaying additional features interpreted as highly spatially variable retrograde zonal and meridional winds, as well as prograde zonal winds. \\   
The most recent results \citep{hueso2008,lellouch2008,clancy2011} and the accumulated evidence for spatial and temporal variations in the upper mesosphere wind field hence support a more complicated picture for the mesospheric circulation than a simple combination of SSAS and RSZ winds.\\
\\
One key factor for improving the knowledge and understanding of the upper mesosphere dynamics is the increase of spatial resolution in mapping mm-wave Doppler-shifts. Currently available single-dish facilities can offer at best a resolution of 8" at 345~GHz and 21" at 115~GHz (with the IRAM-30 m telescope), while Venus' disk size varies between 8" to 58". In addition to smoothing out the information over relatively large spatial scales, single-dish observations typically require multiple pointings to reconstruct the wind field across Venus' disk, at the expense of integration time per pointing. Interferometric mapping of Doppler-shifts obtained with antenna arrays allow for an improved spatial resolution, which is especially necessary to map the disk at maximal elongation and superior conjunction geometries, for which the day-side of Venus is observable. A resolution better than 3" was for example reached in the only previously published Venus' interferometric Doppler-shift maps \citep{shah1991}, obtained at the Owens Valley Radio Observatory (OVRO) array. \\
\\
This paper presents interferometric observations of the CO(1-0) (115~GHz) rotational line obtained at the Institut de Radio-Astronomie Millim\'etrique Plateau de Bure Interferometer (IRAM-PdBI) in November 2007 and June 2009, in 2 consecutive days for each period. All observations were performed near the maximum western elongation of Venus, mapping the morning hemisphere with a spatial resolution between 3.5" and 5.5". We present the obtained Doppler-shifts maps and their possible interpretation in terms of wind field using different combinations of SSAS and zonal winds, along with the determination of the sounded altitudes, to provide the most possible complete snapshot view of the upper mesosphere dynamics on Venus. 

\section{Observations}

\subsection{November 2007}
\begin{figure*}
\begin{center}
\includegraphics[width=10cm]{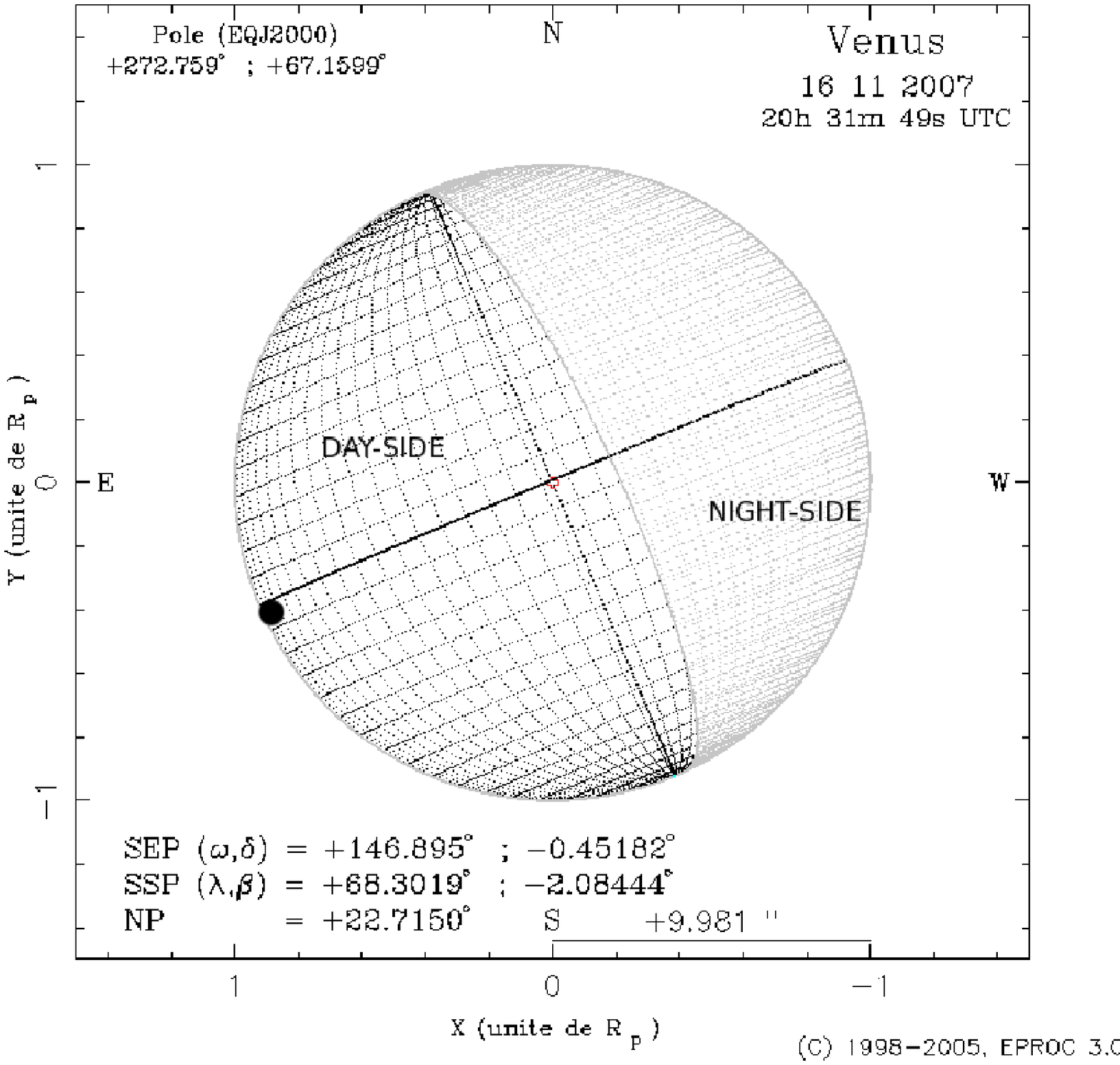}
\includegraphics[width=10cm]{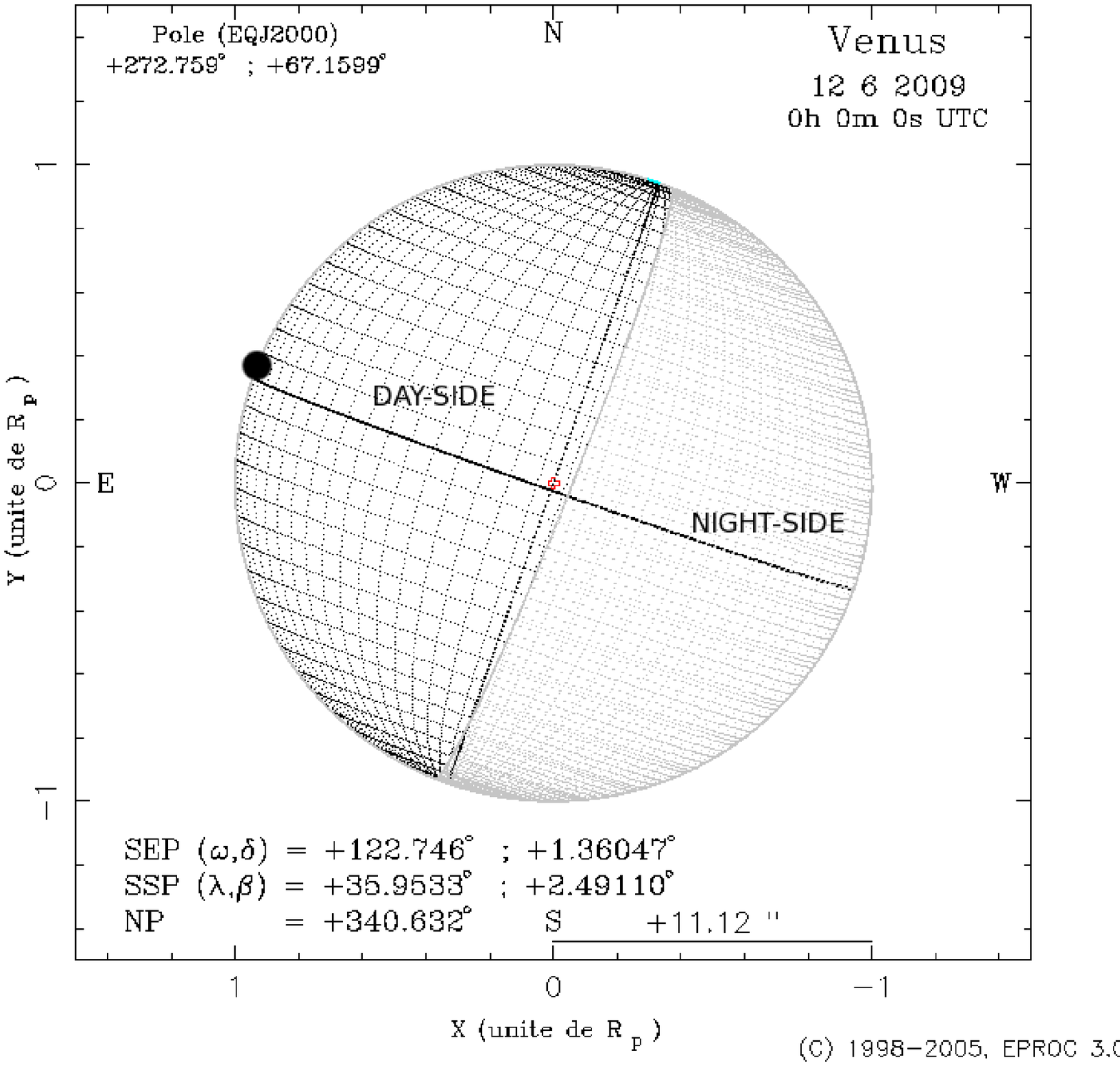}
\end{center}
\caption{\label{imcce} Aspect of Venus on November 16, 2007 (top) and June 12, 2009 (bottom). The black dot next to the eastern limb represents the sub-solar point. The small circle at the center represents the sub-earth point. Coordinates of the sub-earth and sub-solar points are indicated in the East longitude convention, along with the pole axis tilt angle. From www.imcce.fr.}
\end{figure*}

\begin{figure*}
\begin{minipage}{0.5\linewidth}
\includegraphics[width=8cm,angle=-90]{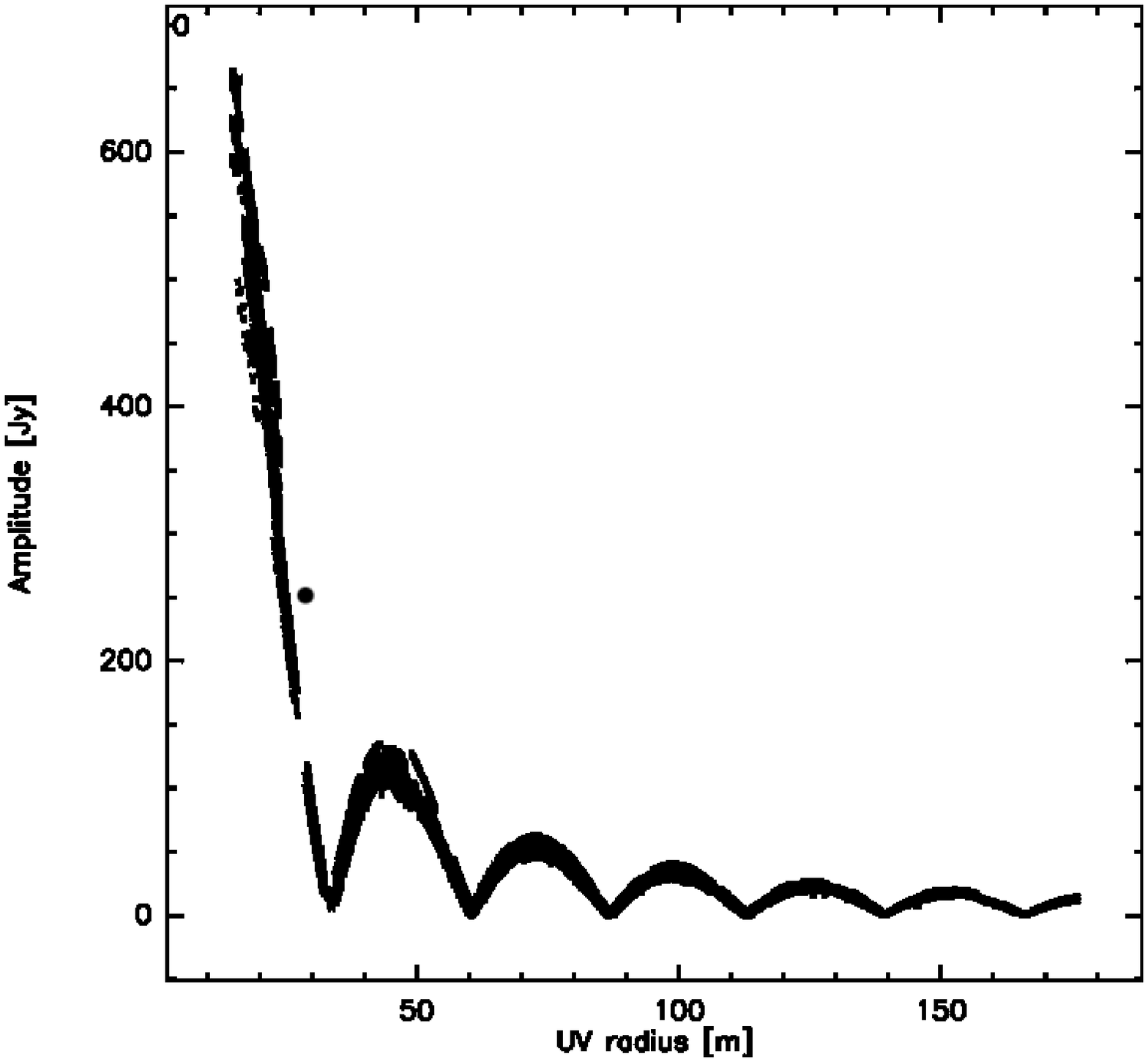}
\end{minipage}
\hspace{0.2cm}
\begin{minipage}{0.5\linewidth}
\includegraphics[width=8cm,angle=-90]{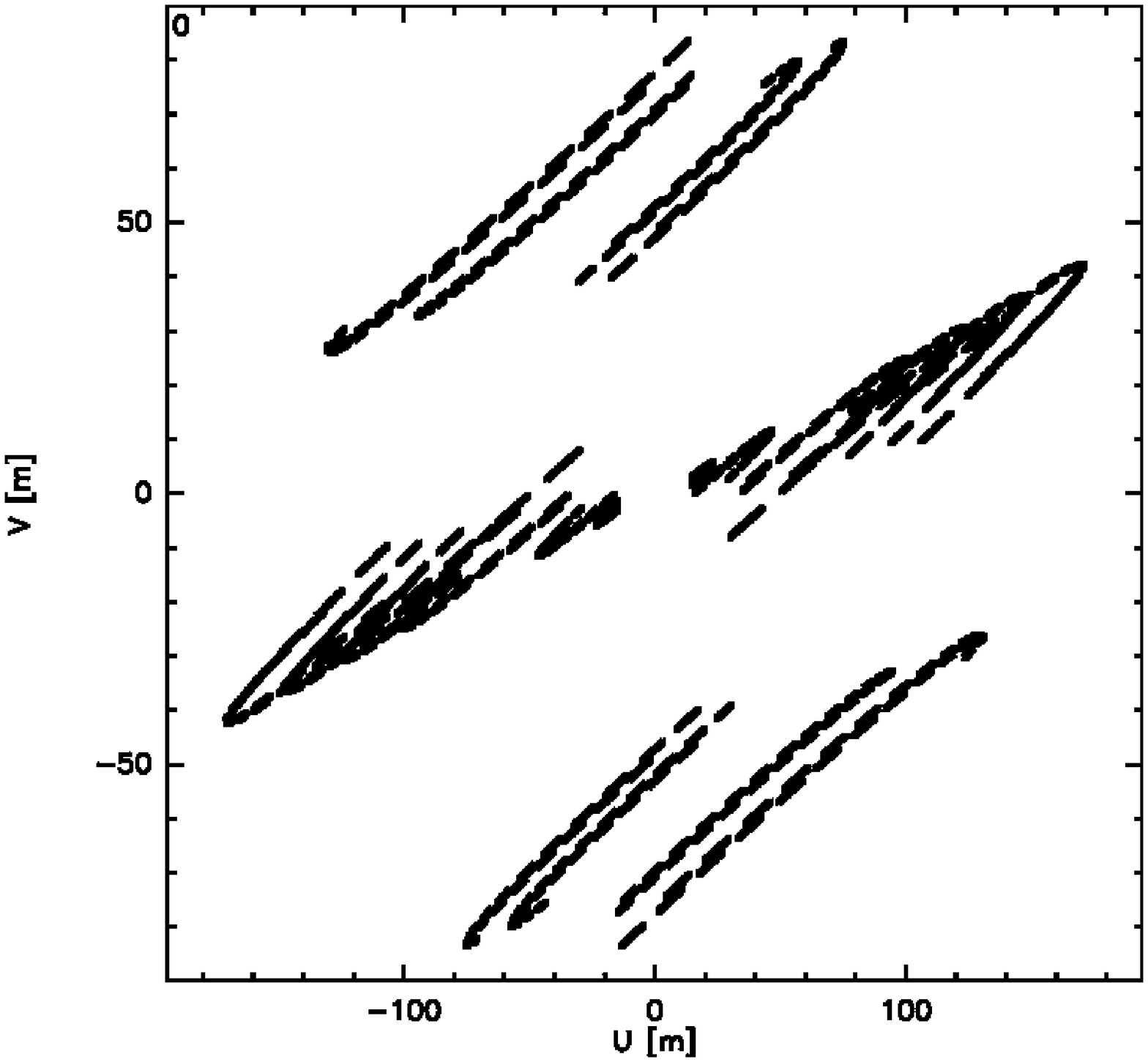}
\end{minipage}
\begin{minipage}{0.5\linewidth}
\includegraphics[width=8cm,angle=-90]{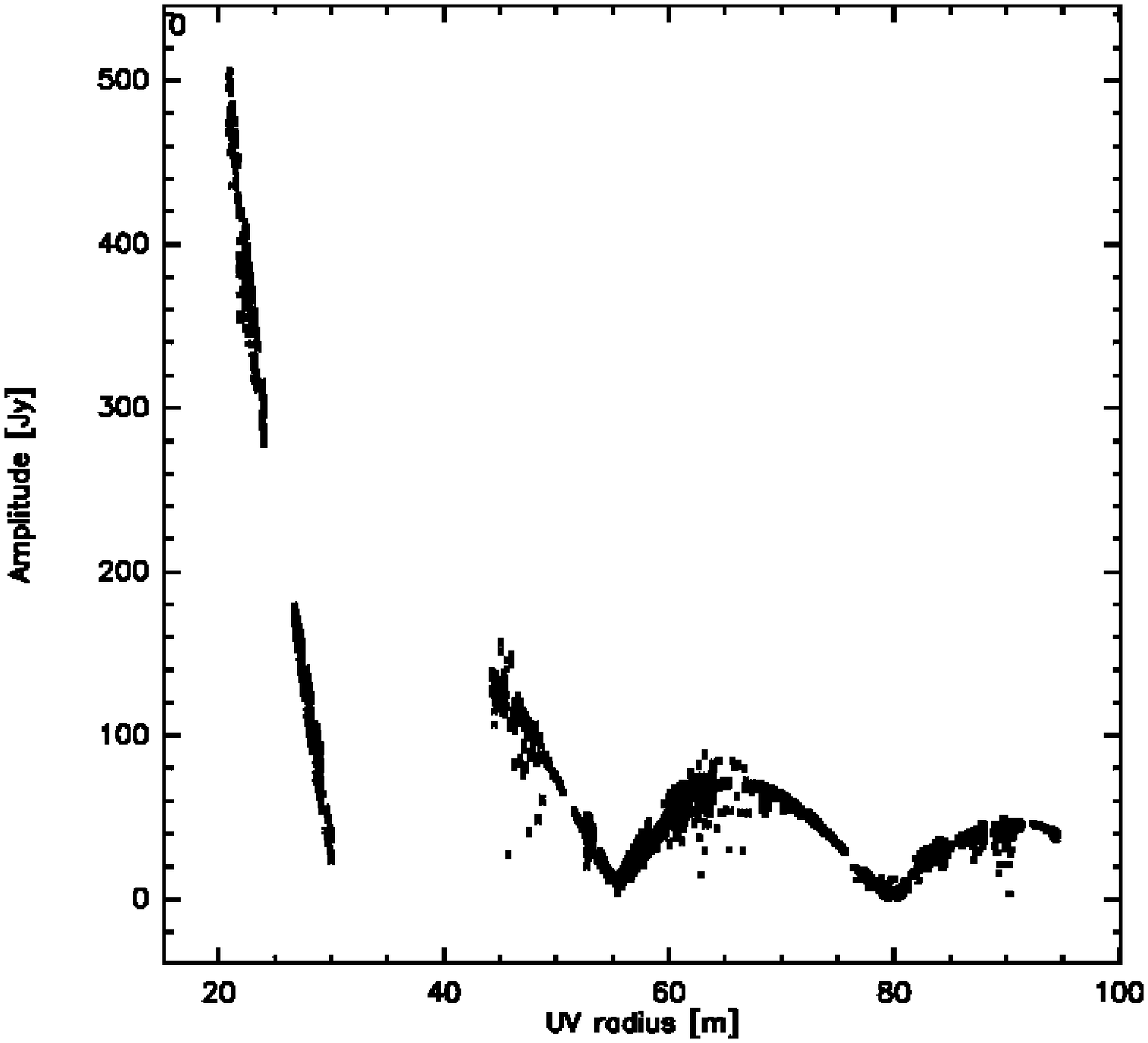}
\end{minipage}
\begin{minipage}{0.5\linewidth}
\hspace{0.2cm}
\includegraphics[width=8cm,angle=-90]{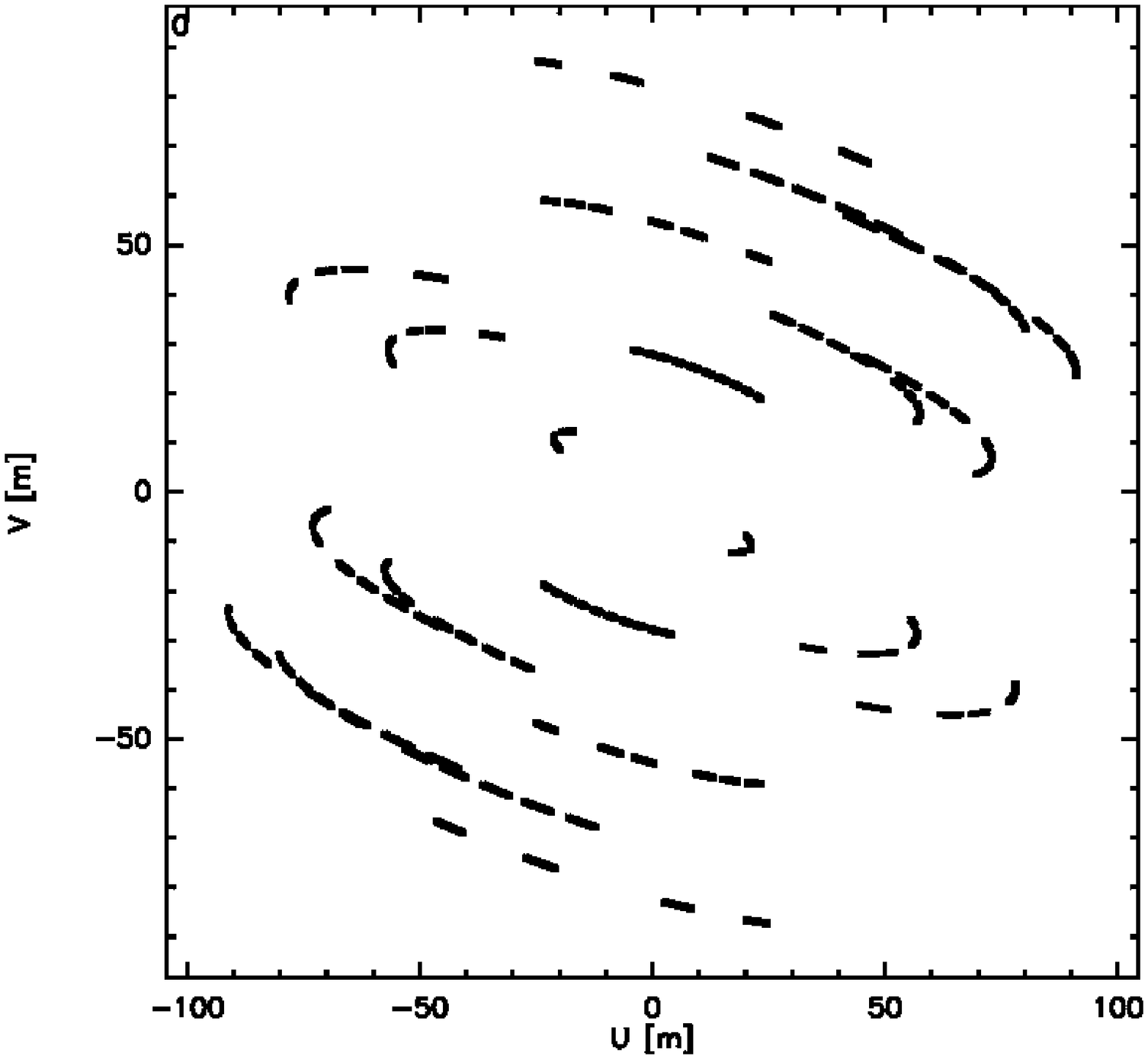}
\end{minipage}
\caption{\label{visib} Continuum visibilities obtained in November 2007 (top panels, merged data from the two observing dates) and in June 2009 (bottom panels, data from June 13). Left panels : amplitude (in Jy) of the visibilities, plotted against projected baseline length (in meters). Right panels : corresponding Fourier-plane coverage. The units correspond to the component of the projected baselines in the East-West (U) and North-South (V) directions (in meters).}
\end{figure*}

The first set of observations was obtained on 16 and 17 November, 2007, with the IRAM-Plateau de Bure Interferometer, located in the Southern French Alps. Observational parameters and ephemeris are gathered in Table \ref{ephem}. At the time, Venus was close to its maximum western elongation, corresponding to the morning hemisphere view, so that both the day-side and the night-side could be observed (see Figure \ref{imcce}, top panel). This geometry is the most appropriate to directly measure the zonal wind contribution near the limbs, where, according to the \citet{bougher1986} model, the contribution of SSAS wind should be minimal.\\
The array, with 5 operating antennas of 15~m diameter each, was in the C configuration with maximum baseline length (distance between antennas projected in a plane perpendicular to the line-of-sight) of 180~m in the East-West direction. The obtained coverage of the Fourier plane, determined by the relative positions of the antennas and the track of the source in the sky, is represented in Figure \ref{visib} (top-right panel). The Fourier-plane filling appears to be very poor, especially in what corresponds to the North-South direction in the image plane, due to the low declination of Venus at the time of the observations ($\sim$$1^{\circ}$).\\
After merging the two datasets, the synthesized beam obtained, that corresponds to the spatial resolution on the final map, is a very elongated ellipse with a half-power beam width (HPBW) of $\sim$5$\times$2" at 115~GHz. Figure \ref{visib} (top-left panel) shows the amplitude of the obtained data (i.e., complex visibilities in the Fourier plane) as a function of projected baseline length. The observed shape roughly corresponds to the Fourier transform of a disk and the multiple nulls evidence the high spatial resolution achieved. The primary beam of the antennas (44.5") is sufficiently large to not resolve Venus' disk ($\sim$20" diameter), nonetheless convolution by the primary beam is taken into account in the models presented in the next sections.\\
The CO(1-0) line at 115.271~GHz was observed in the upper sideband (USB) of the tuned receivers in a 20~MHz-wide spectral window with a 39~kHz spectral resolution (corresponding to a velocity resolution of 101~m/s at 115~GHz), while the continuum was simultaneously integrated on 3 spectral windows totalling 940~MHz of bandwidth. The data were measured simultaneously on 2 cross-polarized receivers which, assuming the source emission is unpolarized, can be considered as equivalent but independent, thus effectively doubling the amount of collected data with respect to a single-receiver system.\\
The calibration of the obtained raw data (visibilities) was performed with the GILDAS-CLIC data reduction package developed at IRAM (http://www.iram.fr/IRAMFR/GILDAS/). Temporal gain calibration of the phase and amplitude of the visibilities was achieved by using as a point-source reference the quasar 1055+018, that was observed every 35 minutes. The atmospheric phase calibration provided by the 22~GHz water vapor radiometers was also applied. The spectral bandpass response was determined using quasars 3c454.3 and 3c273, allowing correction down to the 0.5\% level. \\
To increase the quality of the high spectral resolution data (CO line data), assuming that the continuum distribution is known to a good precision level, the data phase could be self-calibrated using a continuum model (see Section 3.1 for details). The calibrated visibilities could then be imaged by undergoing an inverse fast Fourier transform and deconvolution of the secondary lobes. This was particularly difficult to achieve because of the incompleteness of the Fourier-plane filling. The best results were obtained by combining the two datasets (accounting for the change in the apparent size of Venus by accordingly changing the Fourier-plane scale) and tapering the visibilities, i.e., decreasing the contribution from the longest baselines. After deconvolution using the HOGBOM method \citep{hogbom1974}, the synthesized beam size is 5.7$\times$3.6".\\
The final output of these observations consists of interferometric maps of the continuum emission, and of interferometric maps of the CO(1-0) line (datacubes) obtained from the high spectral resolution window. The noise rms of $\sim$0.7 Jy/beam on each 39~kHz-wide spectral channel is mainly determined by the imaging dynamic range, and corresponds to a signal-to-noise ratio (SNR) of 100 per beam at most.

\subsection{June 2009}
During the June 2009 observations, Venus was in a similar observing geometry as in 2007 (morning hemisphere view, see Figure \ref{imcce}, bottom panel).\\ 
The array was this time in the D configuration (with maximum projected baseline length  of 95~m), and the half-power beam width at 115~GHz was ~4.1$\times$4.1" on June 12 and 5.7$\times$3.4" on June 13, allowing to resolve Venus' disk ($\sim$22" diameter) in up to 6 independent points across.\\
The same spectral setup as in 2007 was used to target the CO(1-0) line.
The phase and amplitude temporal gain calibration was performed in the same manner as in 2007, using as gain calibrators the quasars 0119+115 and 0235+164, and the strong quasar 3c454.3 for bandpass calibration down to a 1\% level. \\
Thanks to a much better Fourier-plane filling than in 2007 (see Figure \ref{visib}, bottom-right panel), the imaging, performed using the SDI  deconvolution method \citep{steer1984}, was much more satisfactory. The June 13 data was eventually tapered, so as to obtain a synthesized beam degraded to a HPBW of 5.7$\times$5.3", ensuring a higher SNR per beam. \\
The final noise rms reached on the interferometric map is of approximately 0.2 Jy/beam on each 39~kHz-wide spectral channel on June 12 (respectively 0.5 Jy/beam on June 13), still mainly limited by the imaging dynamic range. This corresponds to a SNR on the map of 350 per beam at most (resp. 230). 

\section{Data Analysis}

\subsection{Continuum-based phase self-calibration}
In order to obtain satisfactory imaging quality, we choose to self-calibrate the phase of the obtained data visibilities, i.e. to force the continuum data phase to follow a model. This procedure is commonly performed for interferometric datasets \citep[see e.g.][]{moullet2008,moreno2009}, and can significantly improve the phase calibration, provided that the continuum brightness distribution is well known $a~priori$.\\
On Venus, the continuum emission at 115~GHz corresponds to the CO$_2$-CO$_2$ collision-induced pseudo-continuum formed in the cloud deck around 1~bar \citep[40-50~km altitude, see e.g.][]{gurwell1995}. To model this emission, we use a radiative transfer code adapted from \citet{lellouch1994}, assuming standard temperature/pressure profiles, and accounting for curvature at the limb. Minor species SO$_2$ and H$_2$SO$_4$, that produce additional opacity around 115~GHz, are included in the models with a typical content of 130~ppm SO$_2$ and 5~ppm H$_2$SO$_4$ at 40~km. Although some observers have suggested that spatial variations of the continuum level up to 10\% may be present \citep{depater1991}, possibly due to the distribution of absorbers or strong horizontal temperature gradients, we will assume that the continuum emission spatial distribution corresponds to a limb-darkened disk, due to the smooth variation with airmass of the CO$_2$-CO$_2$ opacity. 
The continuum model produced by our radiative transfer code is used as the reference model for phase self-calibration. The corresponding visibilities phases are computed using the same Fourier-plane coverage than the data. Once corrections from the continuum data to the model are derived for each antenna pair (baseline), they are applied to the  phase of the CO line data. We note that the choice of the absorbers content, within a reasonable range, does change slightly the limb-darkening profile of the model, but has very little influence on the result of the self-calibration.\\
The assumed continuum model produces a disk-averaged brightness temperature of 330~K at 115~GHz, that is used as the reference to assess the absolute flux scale of the data.

\subsection{CO line maps}

\begin{figure*}[t]
\begin{center}
\begin{minipage}{10cm}
\includegraphics[width=7.5cm]{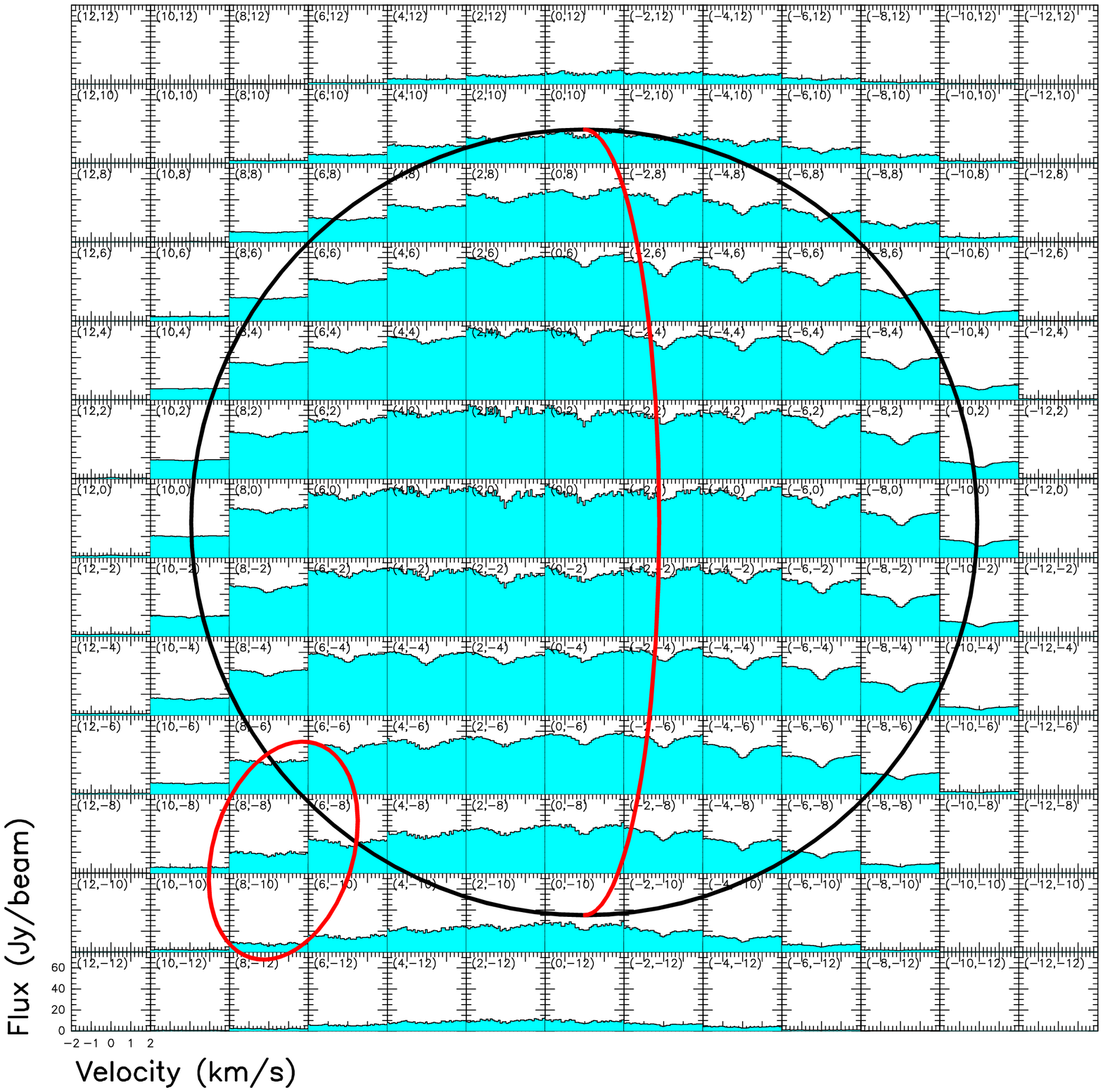}
\includegraphics[width=7.5cm]{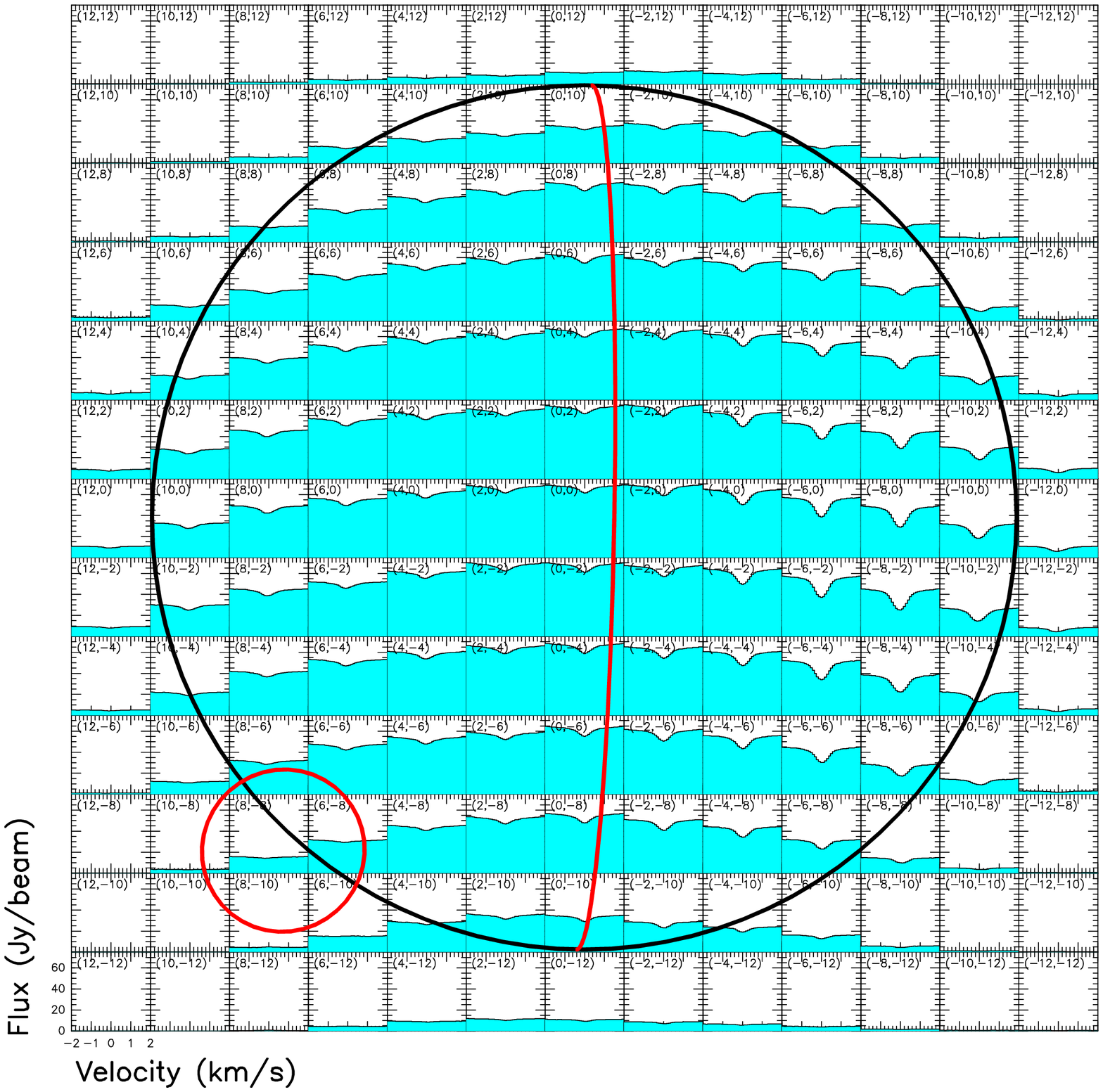}
\includegraphics[width=7.5cm]{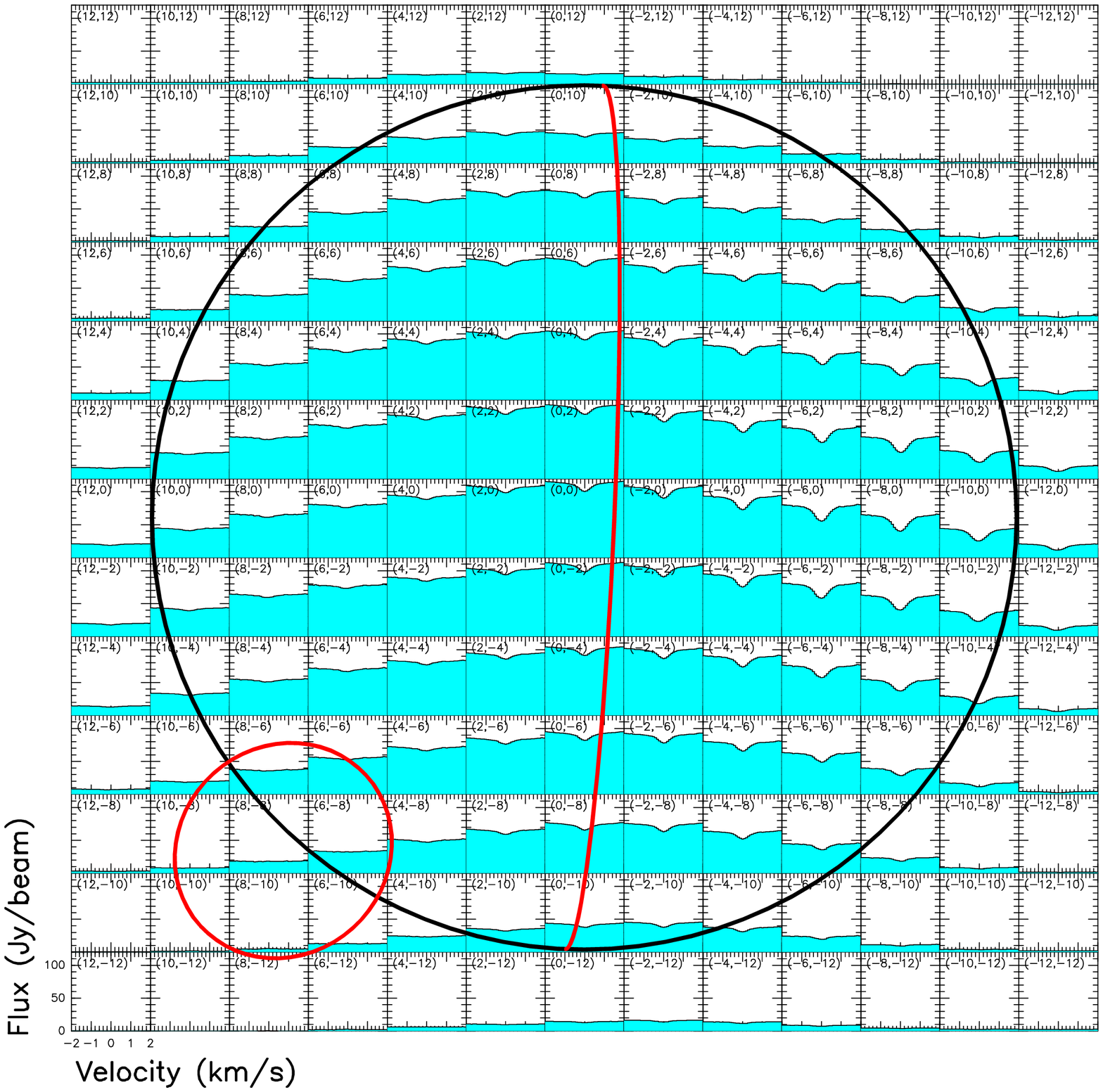}
\end{minipage}
\end{center}
\caption{\label{spectramaps2007} Maps of Venus' CO(1-0) spectra (in Jy/beam), for different observing dates. The location of each spectrum with respect to the disk center is indicated in parenthesis (in arcseconds). The velocity scale for each individual spectrum goes from -2 to +2~km/s. The continuum level is represented by the blue background. In each panel, Venus' disk is represented by the centered circle, the synthesized beam by the red ellipse in the bottom left hand corner, and the terminator by the red vertical arc. The maps were rotated so that the pole axis is vertical in the figure. Top panel : 2007 observations (merged data from the two observing dates). The flux scale goes from 0 to 75~Jy/beam. Middle panel : June 12, 2009 observations. The flux scale goes from 0 to 75~Jy/beam. Bottom panel : June 13, 2009 observations. The flux scale goes from 0 to 120~Jy/beam. }
\end{figure*}

\begin{figure*}
\begin{center}
\includegraphics[width=10cm]{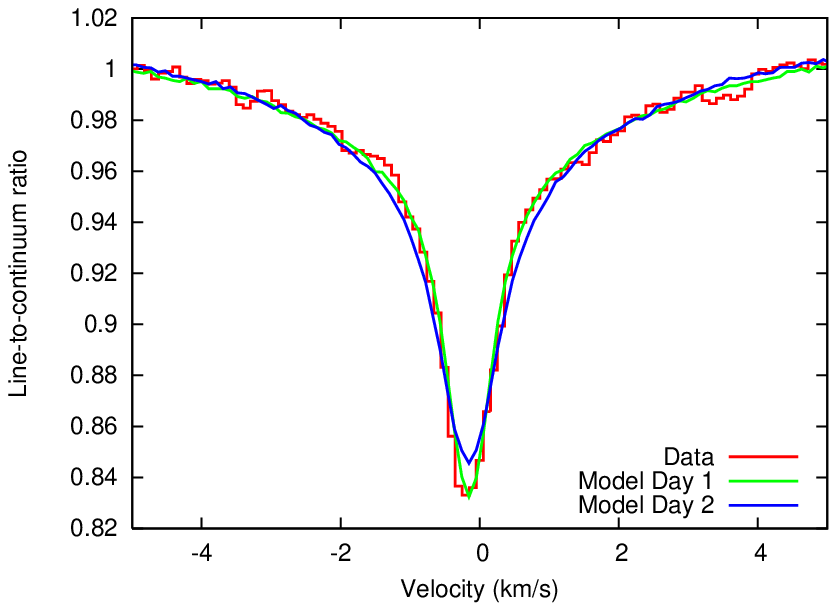}
\includegraphics[width=10cm]{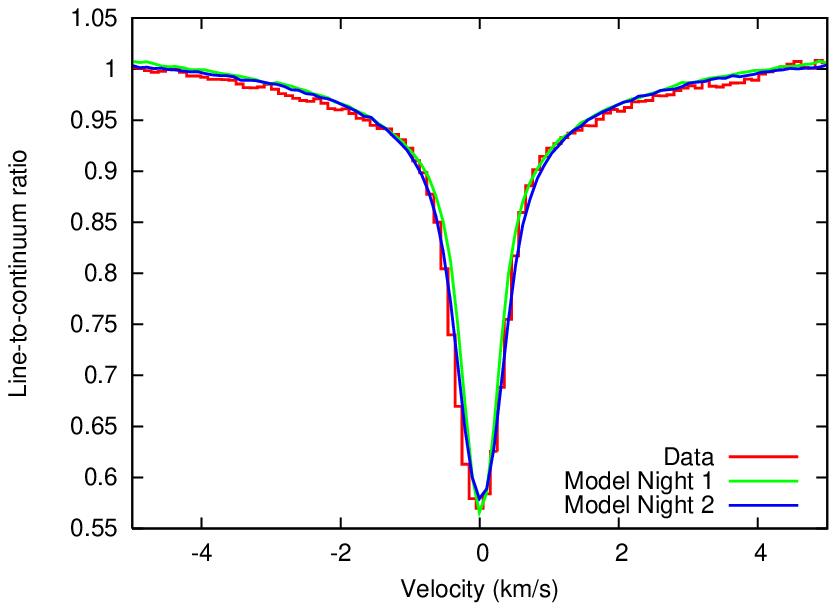}
\end{center}
\caption{\label{fits} CO(1-0) lines measured on June 12,  2009 (in red), plotted against synthetic lines computed assuming adjusted atmospheric models. Model 1 (for day and night conditions) corresponds to a model where only the temperature profile is adjusted, while the CO mixing profile is fixed to the reference profile. Model 2 (for day and night conditions) corresponds to a model where only the CO mixing profile is adjusted, while the temperature profile is fixed to the reference profile. Top : day-side spectra measured on the equator at 8" from the sub-earth point (9.3~am local time). Bottom : night-side spectra measured on the equator at 8" from the sub-earth point (3.1~am local time).}
\end{figure*}

The CO(1-0) line datacubes are represented on Figure \ref{spectramaps2007}  as maps of spectra, sampled on a grid with 2" resolution, where each spectrum corresponds to the local, beam-convolved spectrum. These maps evidence the final imaging quality that could be achieved. All maps display absorption lines all over the disk, with the expected characteristic profile, i.e. lorentzian-shaped broad wings and a Doppler-broadened core of $\sim$1~km/s width. One day-side line and one night-side line are plotted with an expanded scale in Figure \ref{fits}, showing the excellent quality of the individual spectra obtained in 2009.  \\
Since, within the CO line, the brightness temperature at a given frequency is controlled by the atmospheric structure (temperature and CO mixing ratio vertical profiles, and the airmass at Venus) in the sounded atmospheric levels, a qualitative assessment of the variation of the CO(1-0) line-to-continuum ratio across Venus' disk can give an indication on the horizontal distribution of temperature and/or CO mixing ratio in the upper mesosphere. \\
Previous mm-wave observations indicate deeper CO line absorptions on the night-side than on the day-side \citep[see e.g.][]{clancy1985}. The present observations qualitatively confirm those findings. In 2007, the line-to-continuum ratio  on the equator varies from 11\% to 48\% from the sky-East (noon) to sky-West limb (midnight), with a maximum depth near 1~am local time. In 2009, the ratio varies from 19\% to 47\% on June 12, (respectively 12\% to 47\% on  June 13), and the maximum depth is measured near 3~am local time. The variation of the CO line depth has been interpreted as the signature of a steeper thermal gradient on the night-side compared to the day side, and/or an increased mixing ratio of CO  (by a factor up to 4 with respect to the day-side, \citet{clancy1985}), resulting from a strong SSAS wind displacing the CO away from the day-side (where it is produced by CO$_2$ photolysis). With only one line available in our dataset, we cannot distinguish between the two hypotheses.
\\

\subsection{Doppler-shifts maps}
\begin{figure*}
\begin{center}
\begin{minipage}{10cm}
\begin{center}
\includegraphics[width=7.5cm]{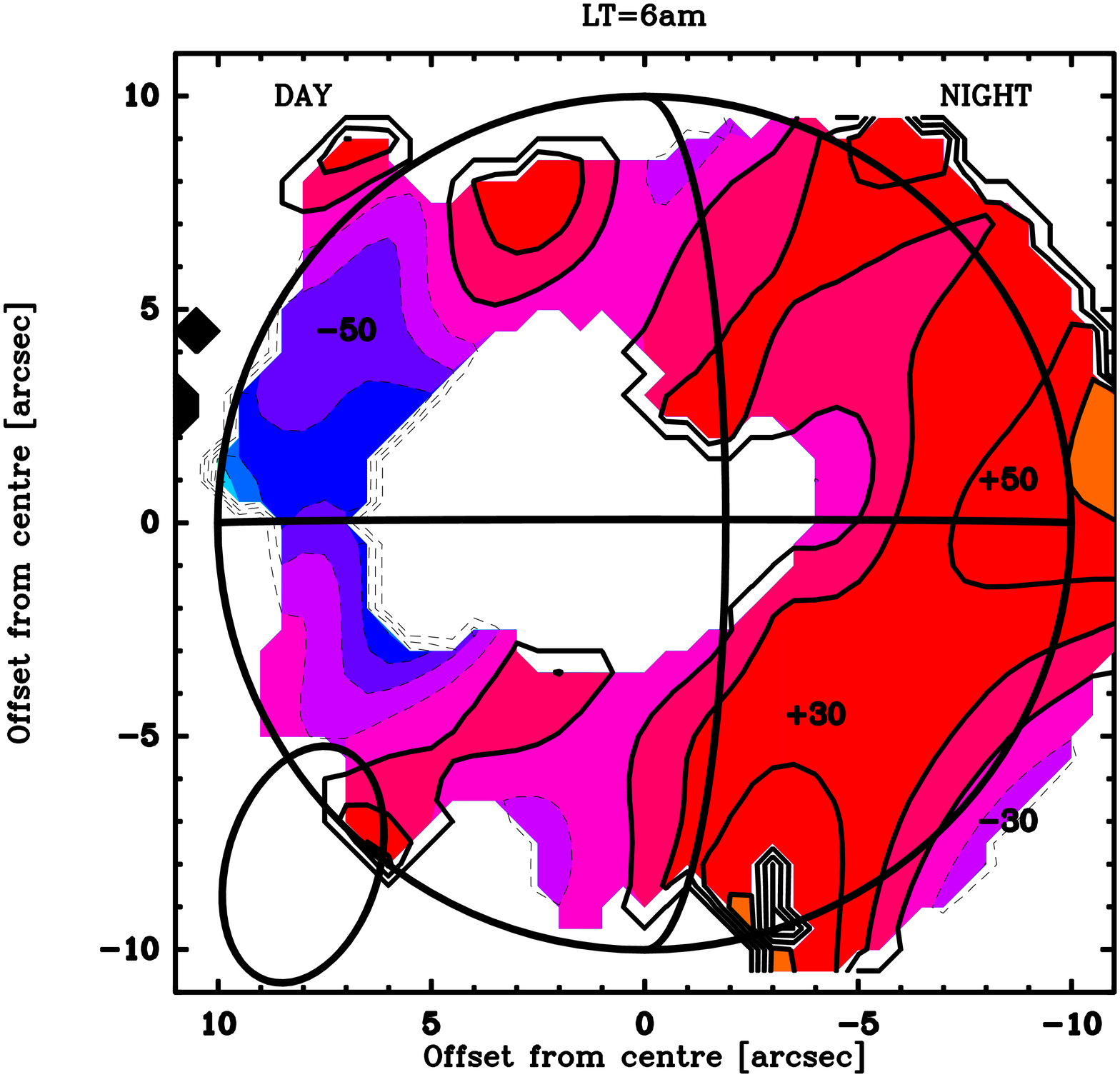}
\includegraphics[width=7.5cm,angle=0]{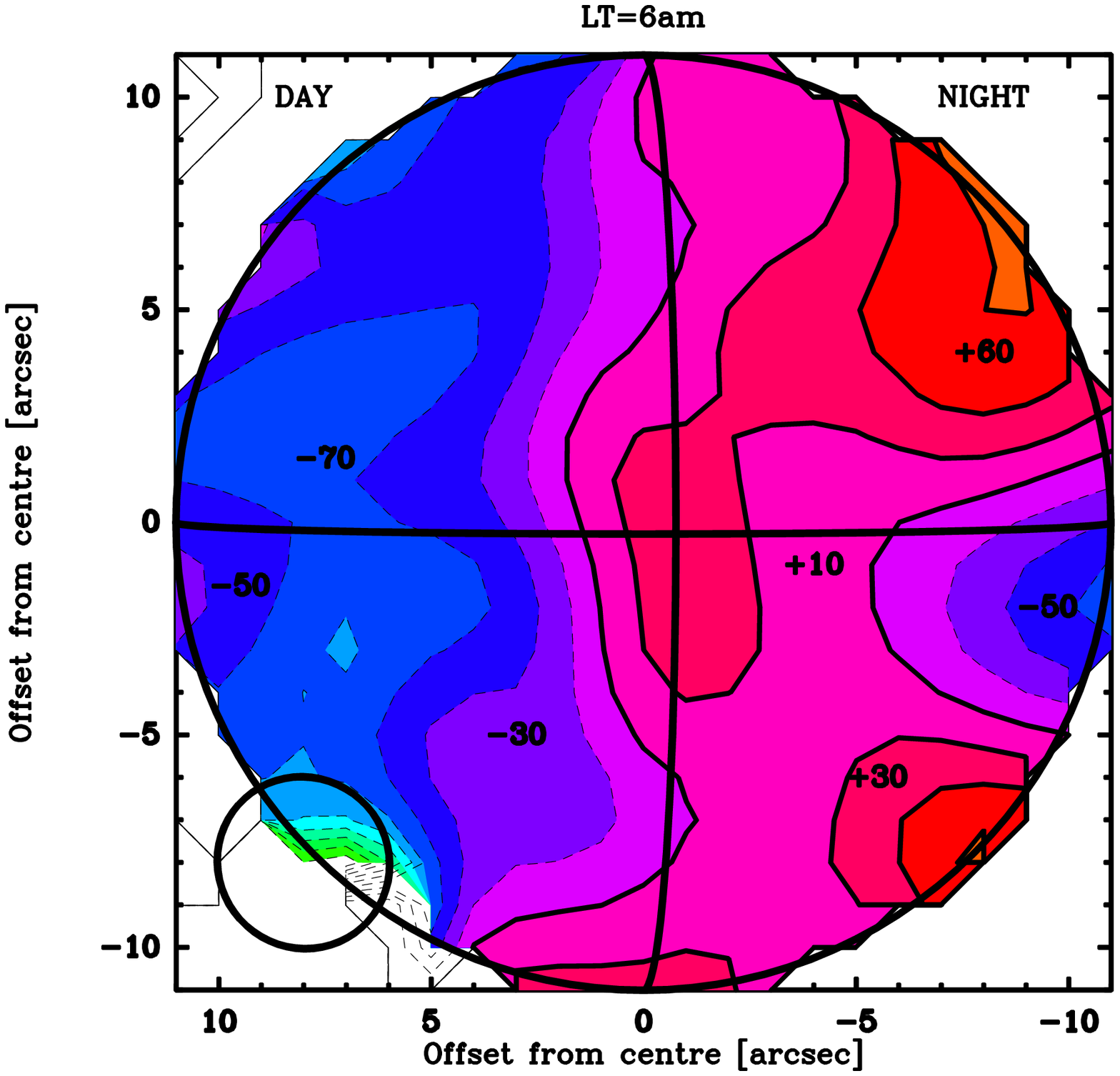}
\includegraphics[width=7.5cm,angle=0]{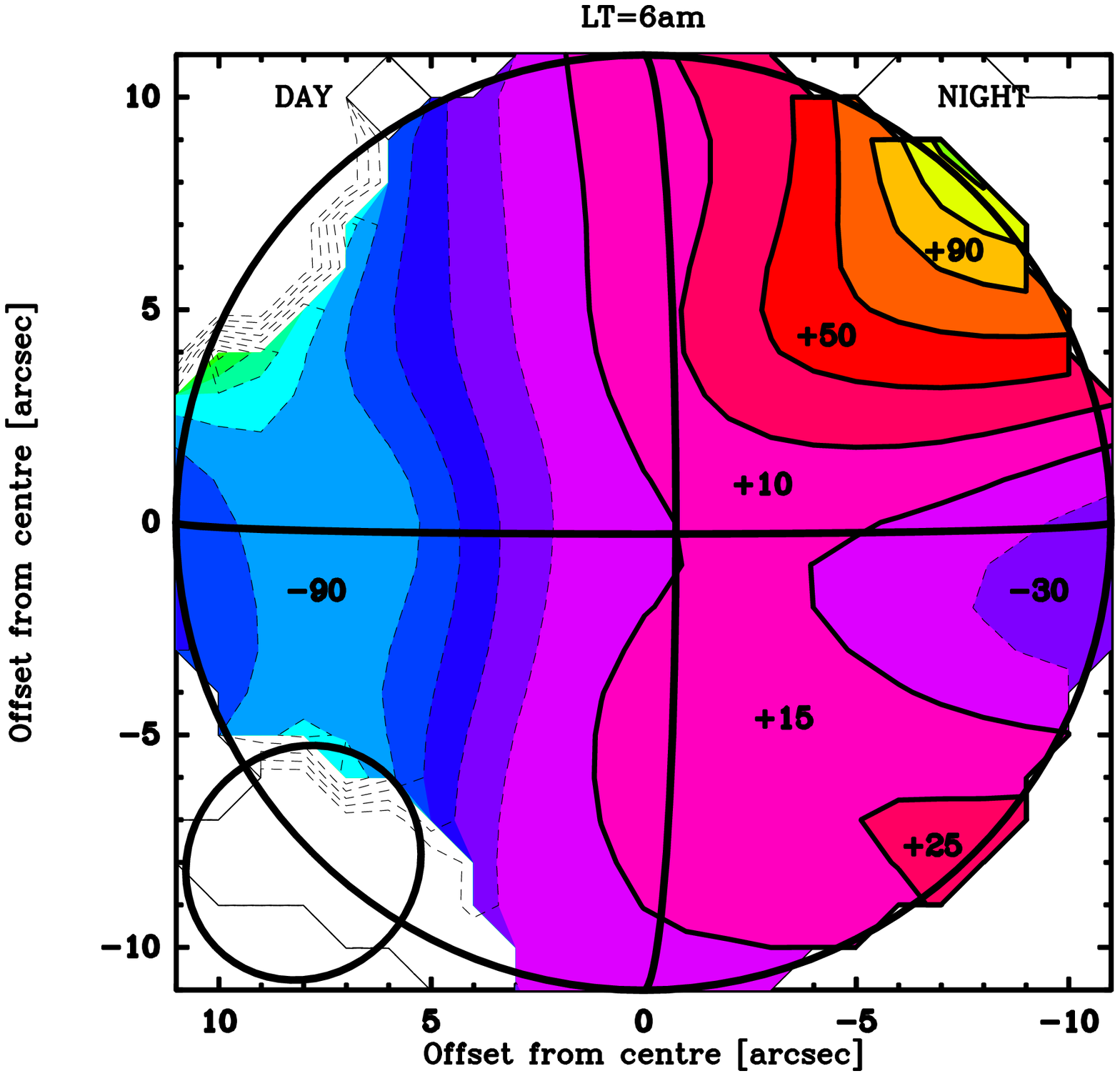}
\end{center}
\end{minipage}
\end{center}
\caption{\label{winds2007} Maps of the Doppler-shifts (in m/s) measured in the core of the CO(1-0) lines, for different observing dates. Regions where Doppler-shifts could not be retrieved appear as blank. In each panel, Venus' disk is represented by the centered circle, the synthesized beam by the ellipse in the bottom left hand corner, and the terminator by the vertical arc. The contour step corresponds to 20~m/s. The maps were rotated so as the pole axis is vertical in the figure.
Top panel : 2007 observations (merged data from the two observing dates). Doppler-shifts errors are of the order of 25-35~m/s on the day-side and 15-20~m/s on the night-side. Middle panel : June 12, 2009 observations. Errors are of the order of 10~m/s on the day-side and 6~m/s on the night-side. Bottom panel : June 13, 2009 observations. Errors are of the order of 15-22~m/s on the day-side and 7~m/s on the night-side. }
\end{figure*}

Measurements of the Doppler-shifts of the CO lines (i.e. offsets from the CO(1-0) line rest frequency) were performed across the spectral maps. Specifically, the data cubes were sampled on a spatial grid with 1 arcsecond wide cells, and each spectrum from each cell was analyzed, using the GILDAS-CLASS line analysis package. To take into account only the inner Doppler-broadened line core in the fit, the outer lorentzian profile was fitted by a 7th order polynomial. The fitted spectral baseline was then removed from the full spectrum, so as to extract the core only ($\pm$0.5~km/s). This procedure does not affect the shape of the line core nor changes its Doppler-shift. The obtained line core was finally fit by a gaussian profile. Even if the line core does not have a perfectly gaussian profile, if its profile is symmetric, the results on the central frequency fit are insensitive to the fitting profile. Only gaussian fits performed on lines detected with a confidence higher than 5 $\sigma$ were considered for the analysis. The Doppler-shifts maps obtained by this technique are represented in Figure \ref{winds2007}. \\
\\
On the 2007 data (Figure \ref{winds2007}, top panel), due to the very low quality of the data imaging, even after merging the datasets from the two consecutively observed days Doppler-shifts cannot be measured on a significant fraction of the disk. In particular, in regions where the CO lines are very shallow and the imaging quality is the worst, CO lines could not be detected with a sufficient confidence. Uncertainties on each measurement are also very high. Globally, we note that blue-shifts (corresponding to approaching winds) are dominant on the day-side, with values of 50~m/s at most and measurement errors of the order of 25-35~m/s. On the night-side measurement errors are lower (15-20~m/s), and mostly red-shifts (corresponding to receding winds) are measured with values up to 50~m/s. At southern latitudes near the western limb ($\sim$1~am local time), the Doppler-shifts display a sign inversion, with blue shifts values of -30~m/s$\pm$15~m/s.  Analyzing separately the data of the two observed days, even fewer Doppler-shifts measurements can be performed, but the Doppler-shifts pattern derived is globally the same, suggesting the stability of the wind large scale characteristics over the 24~hour span between the observations, and all together arguing in favor of the reality of the global wind pattern observed. However, the relatively low significance (2-3$\sigma$) of the individual Doppler-shifts obtained does not justify further analysis or accurate modeling of the 2007 data. \\
\\
On the 2009 data (Figure \ref{winds2007}, middle and bottom panels), the uncertainties obtained are much lower, of the order of 10~m/s on June 12 on the day-side (resp. 15-22~m/s on June 13), and of 6~m/s on the night-side (resp. 7~m/s). A more detailed analysis of the Doppler-shifts maps can then be performed.\\
We  observe that the Doppler-shifts maps on the two consecutive days in 2009 present a very similar pattern, with little variations of the measured values, characterized by :\\
- blue-shifts on the day-side, with velocities up to 70~m/s on June 12 (resp. 90~m/s on June 13).\\
- red-shifts on the night-side with velocities up to 60~m/s on June 12 (resp. 100~m/s on June 13), and significant variations with latitude.\\
- blue-shifts in a small region located on the night-side near the equator and at low southern latitudes, hereafter referred to as the blue-night (BN) region, with values up to -50~m/s  on June 12 (resp. -30~m/s on June 13).\\
We also note that the pattern observed in 2009 globally corresponds to the pattern tentatively observed in 2007, although the similarity is difficult to assess quantitatively. 
\\
\\ 
The measured Doppler-shifts correspond to beam-convolved velocities projected along the line-of-sight (LOS), and differ from the actual atmospheric wind velocities by a projection factor depending on the position on the disk, the flow direction and the beam profile. Therefore a wind field cannot directly be proposed based on the LOS winds but requires deconvolution of the various components.

\subsection{Estimation of probed altitude}
\begin{figure*}[t]
\begin{center}
\begin{minipage}{21cm}
\includegraphics[width=10cm]{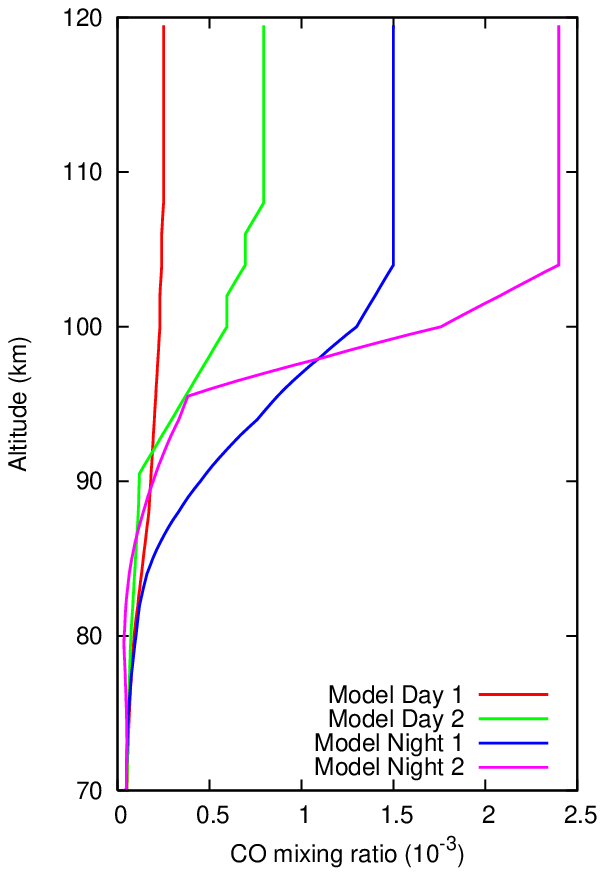}
\includegraphics[width=10cm]{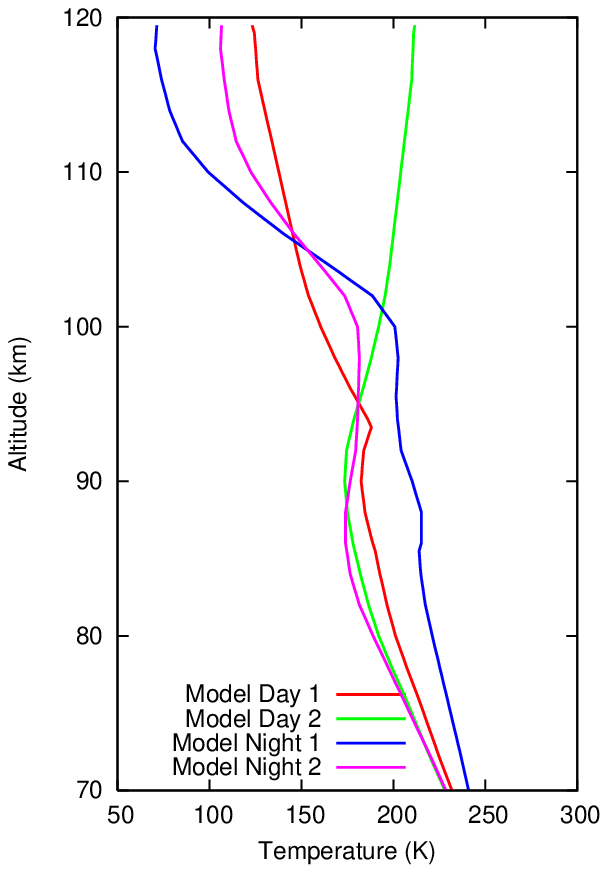}
\end{minipage}
\end{center}
\caption{\label{profiles}  Atmospheric models adjusted on the CO(1-0) lines represented on Figure \ref{fits}. Left : CO mixing ratio profiles. Right : temperature profiles. }
\end{figure*}

\begin{figure*}
\begin{center}
\includegraphics[width=12cm]{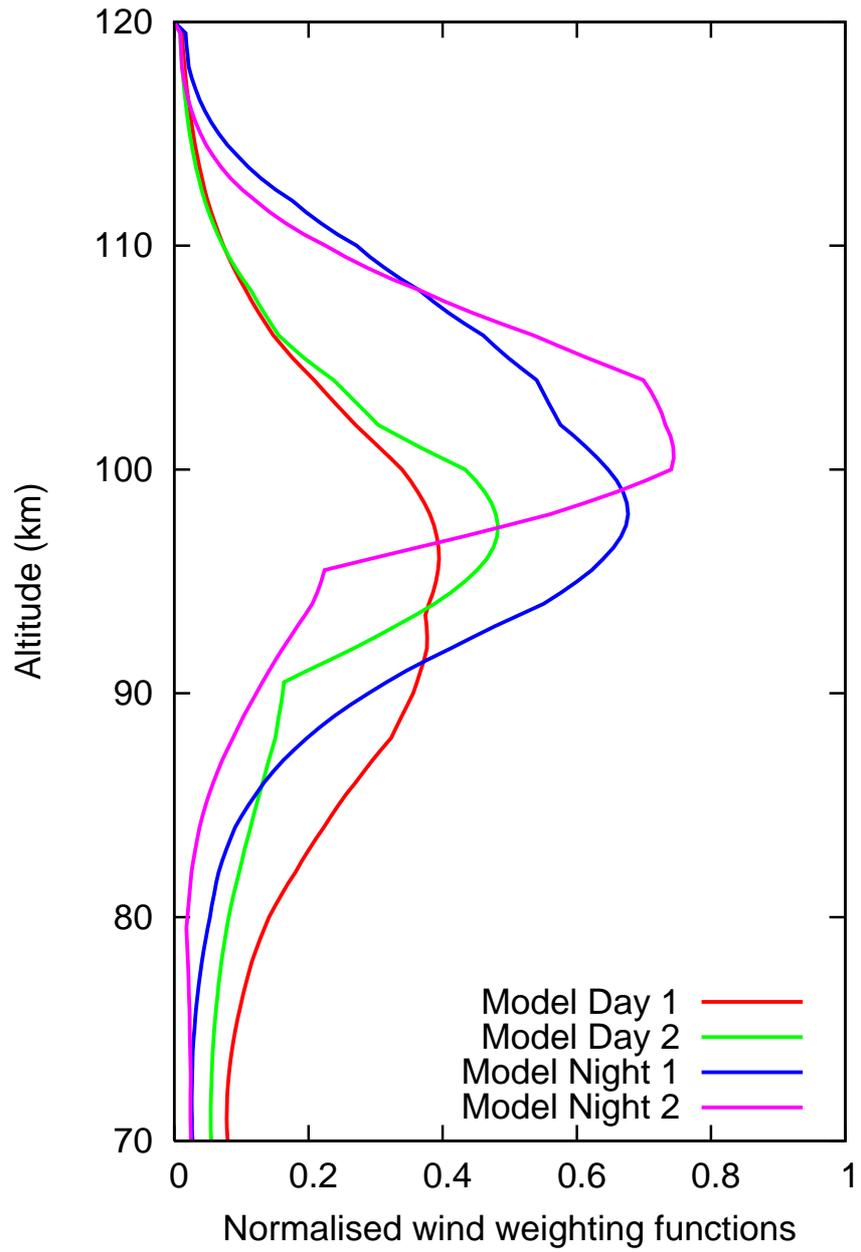}
\end{center}
\caption{\label{wwf} Wind weighting functions computed assuming the adjusted atmospheric models described in Figure \ref{profiles}, for a point on the equator at 8" from the sub-earth point. }
\end{figure*}

To give a sense of the atmospheric region that is sounded by the CO(1-0) line, we performed an approximate modeling of the line emission. The contribution of each atmospheric level is determined by the opacity function of the line, which can be calculated using a radiative transfer model and assuming the atmospheric temperature and CO mixing ratio profiles. When multiple transitions with different opacities are available, it is possible to independently retrieve the temperature and CO mixing ratio profiles from simultaneous fitting of the lines shapes \citep[see e.g][]{clancy2008}, and in turn calculate the contribution from each altitude. However it is impossible to assess unequivocally the temperature and mixing ratio profiles from a single transition dataset. In our case, since we do not know $a priori$ the appropriate temperature and CO mixing ratio profiles, we use an alternate, approximate approach, by considering, for both day and night conditions, two very different atmospheric models that can both reproduce the line profiles, in order to estimate a plausible range for the contribution functions.\\
To determine the two atmospheric profiles considered for this analysis, we use standard profiles for the temperature, pressure and CO mixing ratio for both day and night conditions, taken from \citet{clancy2003}, and modify either the temperature (T) or CO mixing ratio (q) profile to fit the line-to-continuum ratio of the observations. Using a radiative transfer code adapted from the work of \citet{lellouch1994}, and assuming the same continuum model as presented in Section 3.1, we obtain synthetic datacubes that are transformed into synthetic visibilities with the same Fourier-plane coverage as the data, and are then imaged using the same procedure, so as to ensure the best comparison to the observations. \\
We used this method to fit the lines measured on the equator at  9.3~am (day) and 3.1~am (night) local time on June 12, 2009. The synthetic lines obtained from the best models are plotted on Figure \ref{fits}, where the model with modified T profile is referred to as Model 1 and the model with modified q profile as Model 2. The corresponding temperature and CO mixing ratio vertical profiles are presented in Figure \ref{profiles}. We notice that these models reproduce reasonably well the line profiles measured at other positions on the disk, indicating that the atmospheric structure may not change significantly across each (day and night) hemisphere (not shown) .\\
To estimate the altitudes sounded by the Doppler-shifts measured in Section 3.3, we compute wind weighting functions (WWFs), corresponding to the weight functions integrated over the fitted part of the line-core ($\pm$0.5~km/s), weighted by the local slope of the spectrum \citep{lellouch1994}. The WWFs for day and night conditions, represented in Figure \ref{wwf}, show that, at 8" from the sub-earth point, the measured winds probe altitudes between 83-104~km on the day-side and 91-108~km on the night-side, peak respectively near $\sim$96-97~km and 97-101~km, and substantially overlap. At the limb, WWFs peak respectively near $\sim$98~km and $\sim$104~km. Between day and night conditions, the altitude difference of the layer of maximum contribution is thus at most $\sim$6km,  which corresponds approximately to 1.5 scale heights. 

\section{Interpretation of the Doppler-shift field}

We investigate here possible interpretations of the 2009 measurements in terms of wind-field. The Doppler-shifts maps (Figure \ref{winds2007}, middle and bottom panels) show a consistent pattern, that gives indications on the main wind characteristics. The flow appears to globally circulate from  sky-East to West, mimicking a zonal prograde wind, that was seldom observed on Venus and is unexpected from a physical point of view. Given the observing geometry, this could also correspond to a circulation from the day-side to the night-side, i.e. a SSAS wind. In the BN region, the flow direction is inverted, and corresponds to that of a retrograde zonal wind (RSZ). Along the central meridian longitude, the measured Doppler-shifts are either very low or not significant, which argues for the absence of  a measurable meridional wind.  We choose then to only explore combinations of zonal and SSAS winds, using a single wind-field for the entire sounded altitude range. This approximate approach is motivated by the relatively low variation of the sounded altitude range across the disk (see Figure \ref{wwf}). We will not try to adjust a model for each individual Doppler-shift measurement but rather to build a global wind-field model that can reproduce the observed Doppler-shifts distribution as a whole. \\
To compute synthetic Doppler-shifts  maps for comparison to  the observed data, we proceed as follows : wind models  are added to the radiative transfer model, by applying a Doppler-shift (corresponding to the local line-of-sight projected wind) to the opacity spectra computed at every position on Venus' disk for each altitude level. The planet's own solid rotation (1.8~m/s at the equator) is neglected. The obtained models  are transformed into synthetic visibilities with the same Fourier-plane coverage as the data. The models are then imaged and the line cores fitted using the procedure described in Section 3.3.  In this manner, the synthetic models are produced and analyzed exactly as the data.\\

\subsection{Zonal winds model}
Zonal winds are characterized as a longitudinal circulation at constant velocity for a given latitude, that can display variations with latitude and altitude.\\
We implemented a zonal prograde wind model, assuming that the wind mimics a solid rotation, i.e. with a velocity maximal at the equator and varying as cos(latitude), and no variation with altitude. We determine that a 69$\pm$14~m/s equatorial velocity for June 12 (75$\pm$20~m/s respectively for June 13) can approximately match the observed limb-to-limb Doppler-shift difference, with the exception of the blue-night (BN) region (see Figure \ref{combs}, top-left panel). As discussed in Section 3.3, the 2007 data quality is not sufficient to perform this analysis with the same confidence, but we notice that the 2007 Doppler-shift pattern can be best reproduced with a similar prograde zonal wind speed ($\sim$85~m/s).\\
In the context of a zonal wind model, the approaching winds seen in the BN region can be interpreted as the signature of a localized {\it{retrograde}} zonal wind in an equatorial band between approximately 10$^{\circ}$ and -20$^{\circ}$ of latitude. The only possible explanation for this feature would be a change in sounded altitude between the day-side and the night-side, that would also correspond to a change of zonal wind direction.  However, the altitude sounding estimates discussed in Section 3.4 show that the WWF for day and night conditions substantially overlap, and display a difference of 5~km at most for the maximal contribution layer. A radical change of the zonal wind direction in just over a scale height is unlikely, and thus a combination of purely zonal winds is not a favored hypothesis to  explain the flow inversion in the BN region.\\
Even away from the BN region, the adjusted prograde zonal wind models are not entirely satisfactory. In these models the increase of Doppler-shift with longitude from sky-East to West is only due to the change in the projection factor, but the variation observed on the data in the day-side is  more abrupt than the models near the central longitudes and shallower near the limbs. This indicates that the wind velocity on the day-side is not constant for a given latitude, but most likely increases from the limb to the central meridian (i.e. from the sub-solar region to the terminator). This argument further confirms that zonal winds alone are not appropriate to describe the observed wind field.

\subsection{Subsolar to antisolar wind : model from \citet{bougher1986}}

\begin{figure*}

\begin{minipage}{0.5\linewidth}
\includegraphics[width=8cm]{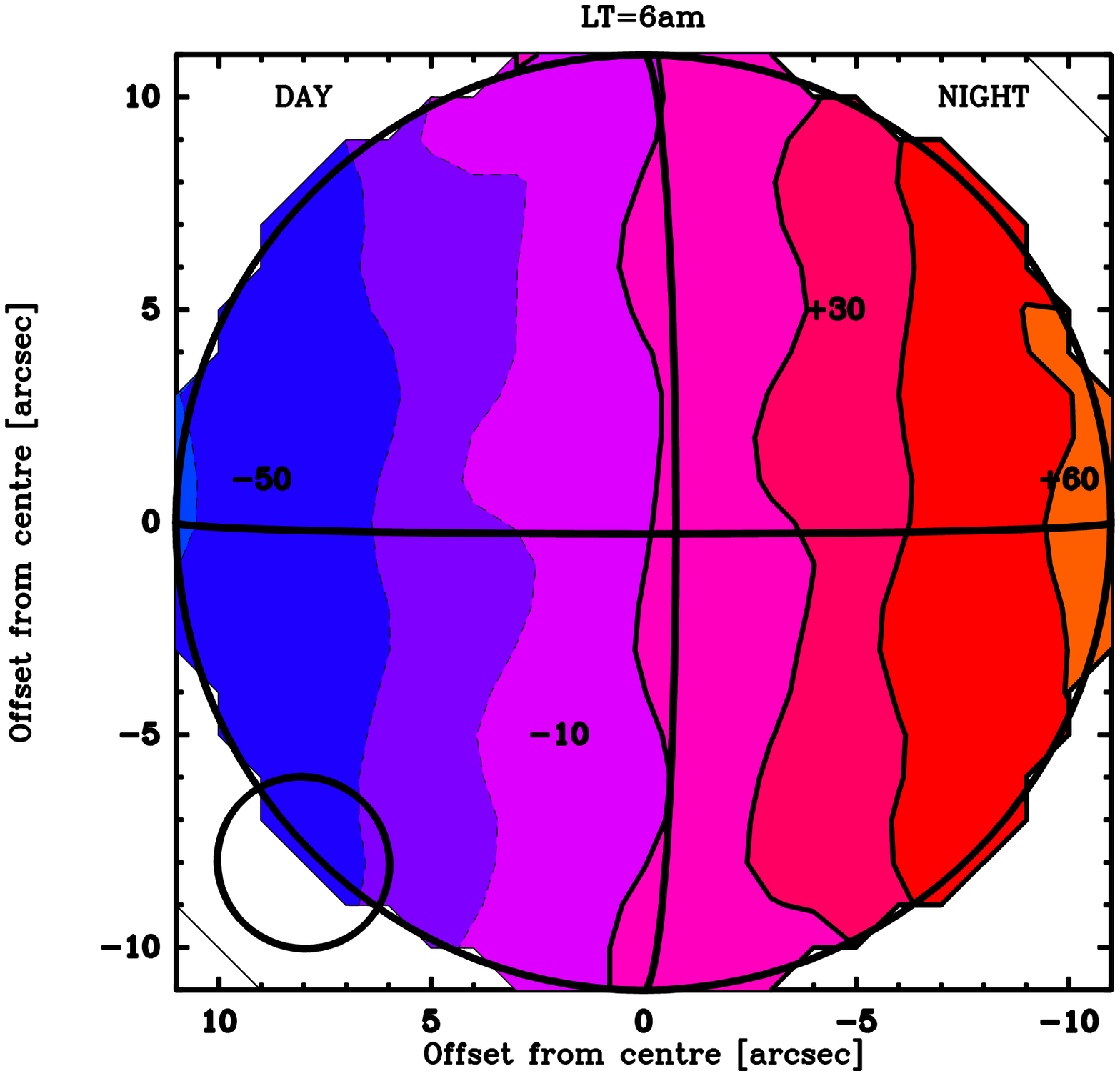}
\end{minipage}
\hspace{0.4cm}
\begin{minipage}{0.5\linewidth}
\includegraphics[width=8cm]{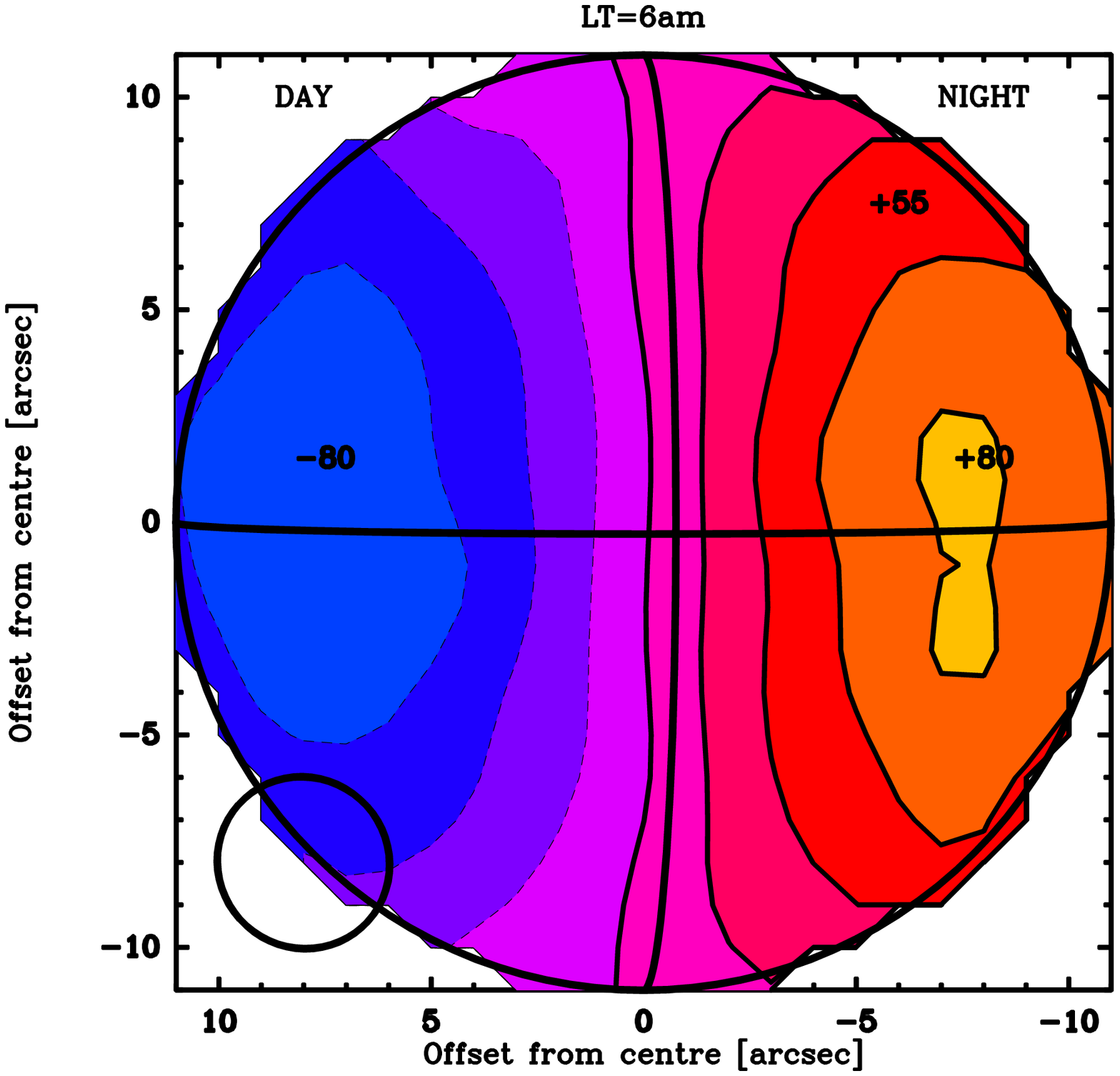}
\end{minipage}
\vspace{1cm}
\begin{minipage}{0.5\linewidth}
\includegraphics[width=8cm]{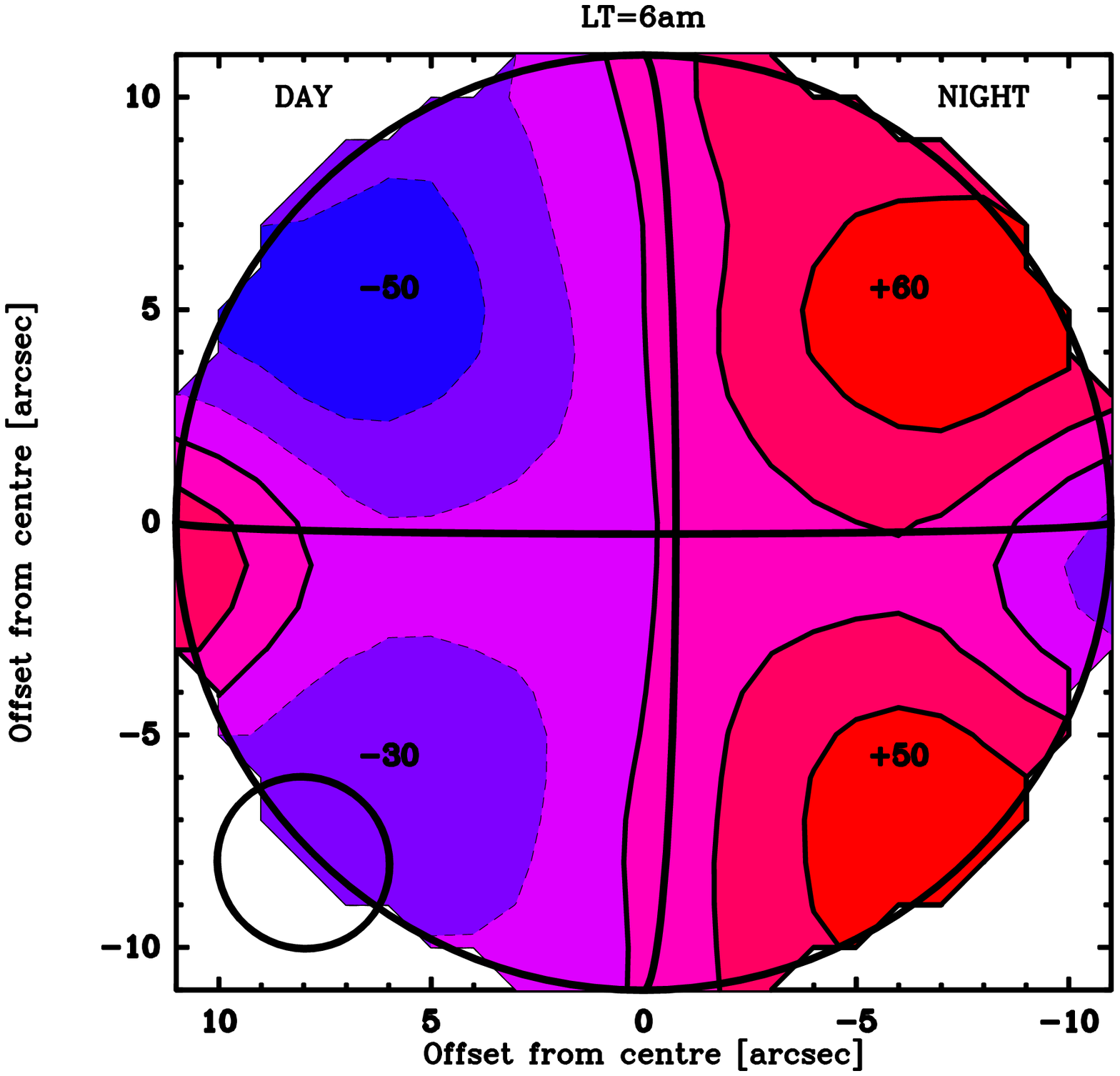}
\end{minipage}
\begin{minipage}{0.5\linewidth}
\hspace{0.4cm}
\includegraphics[width=8cm]{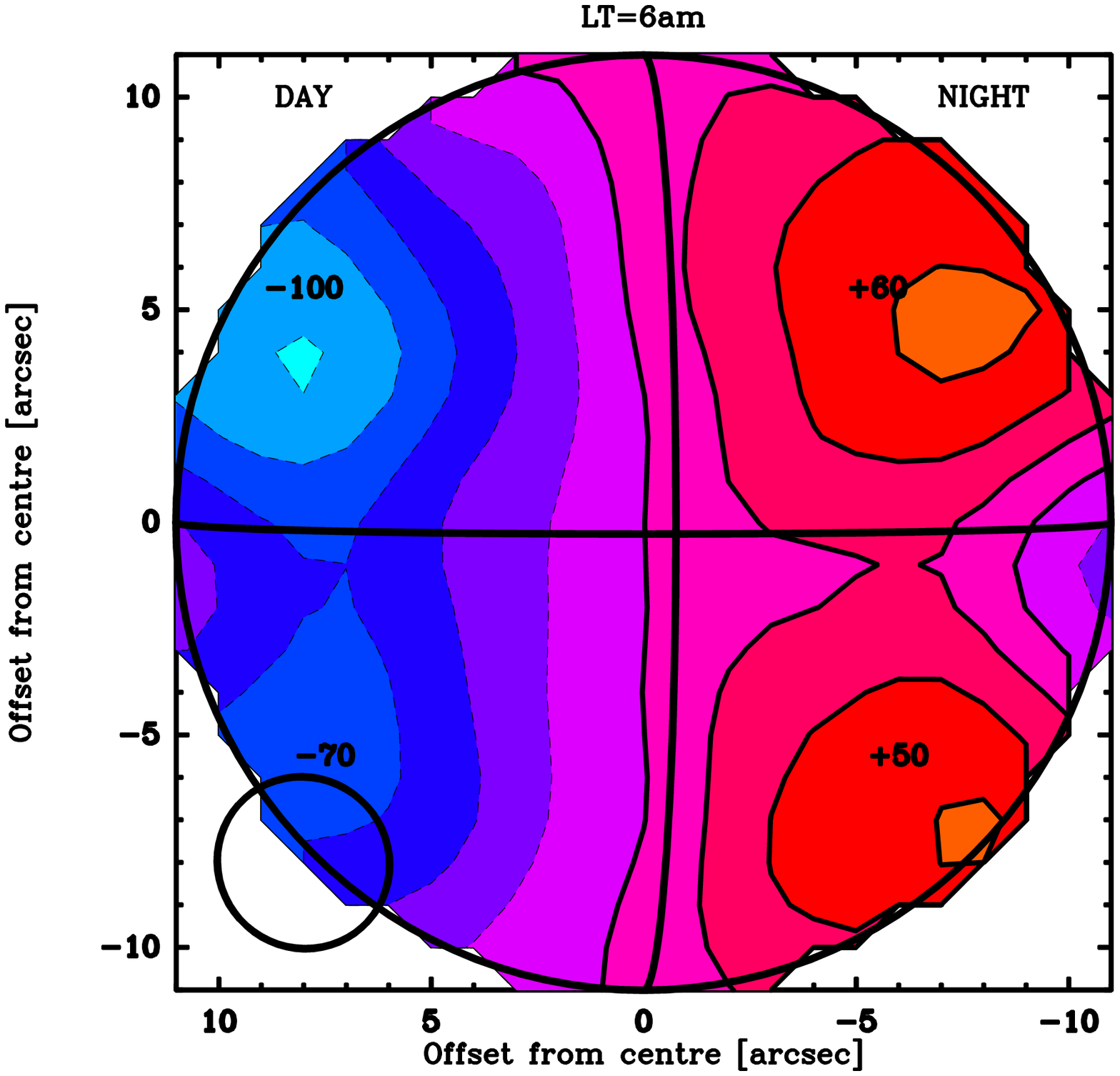}
\end{minipage}
\caption{\label{combs}   Maps of synthetic Doppler-shifts (in m/s), computed assuming different wind models. In each panel, Venus' disk is represented by the centered circle, the synthesized beam by the ellipse in the bottom left hand corner, and the terminator by the vertical arc. The contour step corresponds to 20~m/s. The Fourier-plane coverage of June 12, 2009, was used to simulate the observations.
Top-left panel: prograde zonal wind model with V$_{eq}$=69~m/s. Top-right panel: \citet{bougher1986} SSAS wind model with V$_{ter}$=200~m/s. 
Bottom-left panel: \citet{bougher1986} SSAS wind model with V$_{ter}$=200~m/s, combined with a V$_{eq}$=100~m/s RSZ wind, localized in an equatorial band. Bottom-right panel: same model as in bottom-left panel, but with a modified SSAS wind velocity field in the day-side varying as V=V$_{ter}$*(1-$\big[\frac{\left|(90-sza)\right|}{90}\big]^5)$, where sza is the the local solar zenith angle (solar incidence). }
\end{figure*}

The thermospheric SSAS wind has been modeled by \citet{bougher1986} as a flow originating from the sub-solar point and following radial paths symmetric with respect to the sub-solar to anti-solar line. The corresponding wind field is hence composed of both zonal and non-zonal components. The wind velocity, expressed as V=V$_{ter}$*(1-$\big[\frac{\left|(90-sza)\right|}{90}\big])$,  increases linearly with solar zenith angle (sza), being maximal at the terminator with velocity V$_{ter}$. This model is commonly used as a tool to interpret mm-wave single dish Doppler-shift data  \citep[see e.g.][]{lellouch2008,sornig2011}. \\
We implemented this model taking into account the projection of the wind on the line of sight as presented in  \citet{shah1991} and \citet{goldstein1991}, and assuming no variations with altitude. If, as the model proposes, the SSAS wind velocity at the sub-solar and anti-solar longitudes is null, then the significant Doppler-shifts measured at the limbs in these observations can only be explained by the combined effect of spatial beam convolution and a significant SSAS wind velocity at low solar zenith angle (solar incidence), implying very high maximal velocity at the terminator. By adjusting the maximal velocity, we find that assuming V$_{ter}$=200$\pm$20~m/s for June 12 and V$_{ter}$=225$\pm$25~m/s for June 13, SSAS wind models can globally  better reproduce the Doppler-shift maps than zonal winds models (see Figure \ref{combs}, top-right panel). They however do not produce the flow inversion observed in the BN region. This feature could be interpreted as the signature of the return branch of the thermospheric SSAS wind that is expected to be present at lower altitudes than the SSAS flow, but that was never unambiguously  identified in previous observations. However, as shown in Figure \ref{wwf}, the CO lines on the night-side tend to sound higher altitudes than on the day-side, which argues against this hypothesis. The flow inversion can hence not be explained in the context of a purely SSAS wind field.\\
 Elsewhere, the match between the datasets and the models is not perfect either. In particular, the \citet{bougher1986} SSAS model predicts a North-South and day-night symmetry of the wind-field, which is not observed. The best model presented in Figure \ref{combs} (top-right panel) indeed underestimates velocities measured on the day-side at Southern latitudes (40$^{\circ}$), and overestimates Doppler-shifts on the night-side at Southern latitudes (40$^{\circ}$), by a factor up to four.\\ 
A SSAS wind is then mostly supported by the observations, but the multiple discrepancies between the model and the datasets show that the model proposed by \citet{bougher1986} is, by itself, insufficient.

\subsection{Descriptive model : SSAS wind and RSZ localized wind}

\begin{figure*}
\begin{center}
\includegraphics[width=20cm]{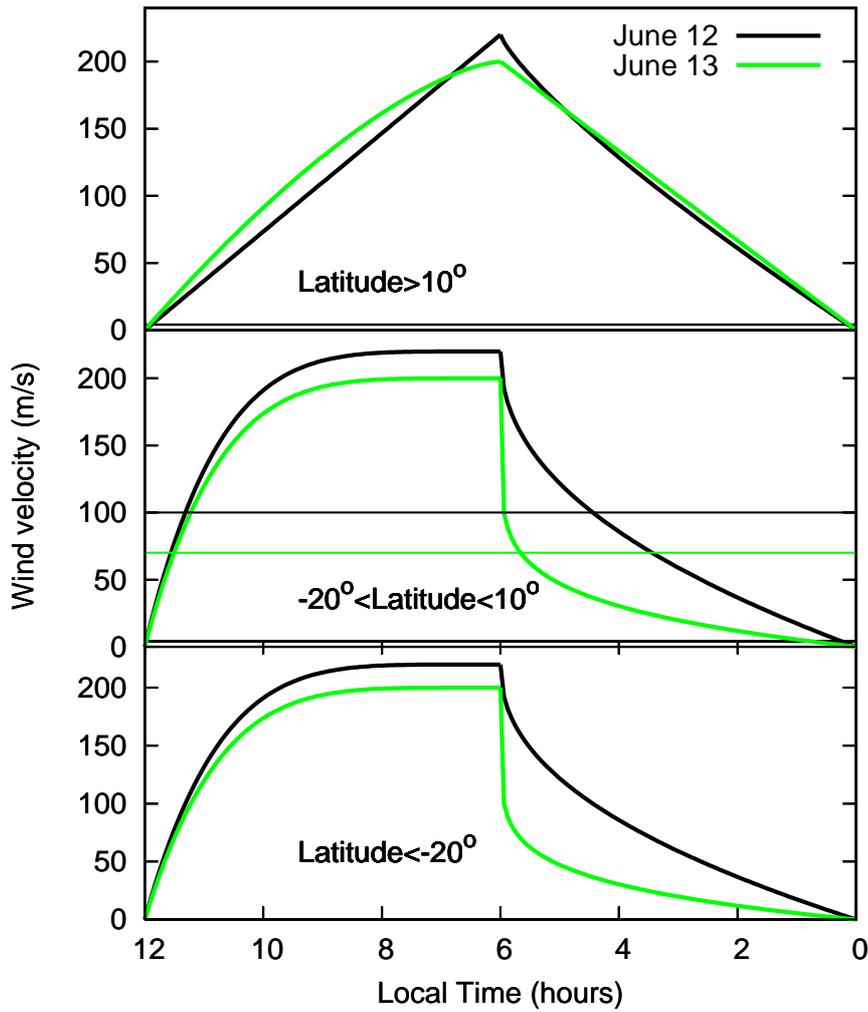}
\end{center}
\caption{\label{proposedssas} Wind velocity as a function of local time for the best fitting parametrized wind models (see Section 4.3 for details), for different latitudinal regions. The thick lines represent the SSAS wind velocity (plotted for an equivalent equator latitude), and the thin lines represent the RSZ wind velocity. Upper panel: models for latitudes $>$10$^{\circ}$. Middle panel: models for latitudes between -20$^{\circ}$ and 10$^{\circ}$. Lower panel: models for latitudes $<$-20$^{\circ}$. }
\end{figure*}

\begin{figure*}
\begin{center}
\includegraphics[width=10cm]{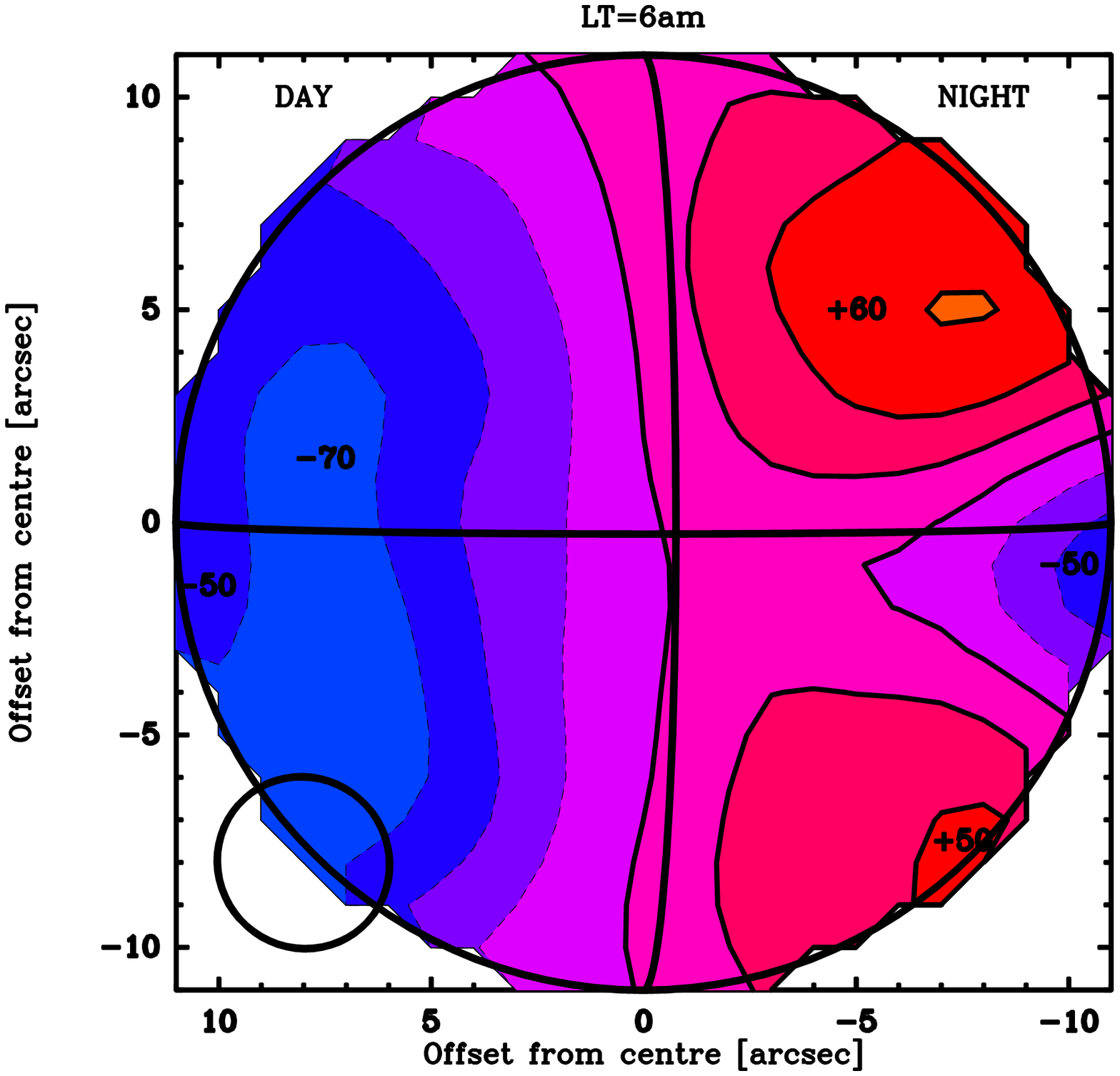}
\includegraphics[width=10cm]{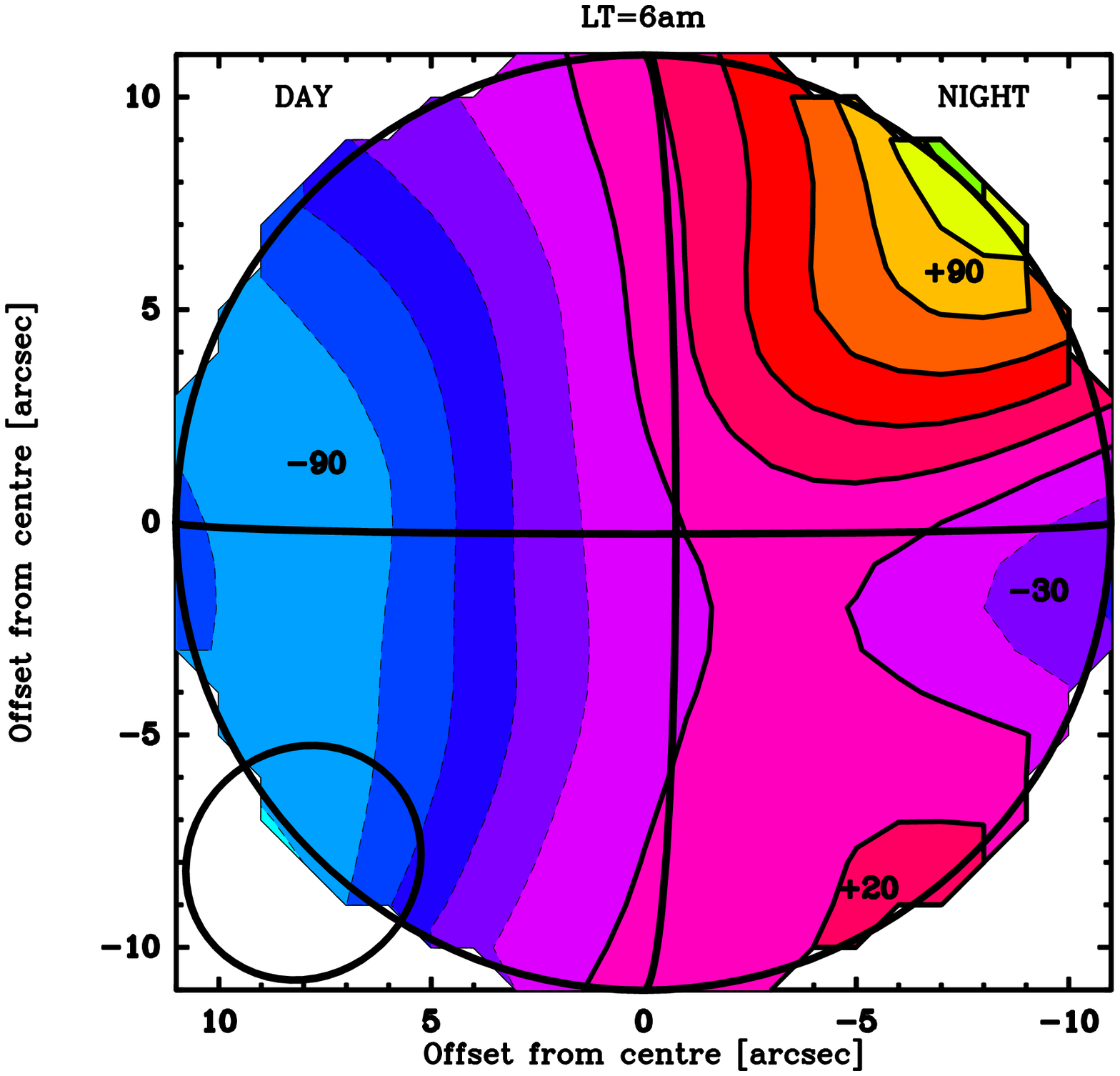}
\caption{\label{adjustedmaps}  Maps of synthetic Doppler-shifts (in m/s) computed assuming the best fitting parametrized wind models, combining a modified SSAS wind-field and a localized RSZ wind. See Section 4.3 for a detailed description of these models. Top : model for June 12, 2009. Bottom : model for June 13, 2009.}
\end{center}
\end{figure*}

In the two previous subsections, using standard wind models to compare with the data, we inferred the following indications on the observed wind-field :\\
- in the BN region, the flow inversion  can be explained by a local RSZ wind, \\
- the observations globally support a SSAS wind, but also indicate significant day-night and North-South asymmetries. 
\\
We explore hereafter combinations of RSZ localized winds with different SSAS wind models, assuming no variations with altitude. \\ 
\\
Figure \ref{combs} (bottom-left panel) shows the example of a simple combination of a SSAS wind (V$_{ter}$=200~m/s, following the \citet{bougher1986} model),  with a RSZ wind (V$_{eq}$=100~m/s) limited to an equatorial band between -20$^{\circ}$ and 10$^{\circ}$ of latitude. This model succeeds in reproducing the wind inversion in the BN region, but, as the model is symmetric with respect to the terminator, also displays a wind inversion feature near the sub-solar region, that is at odds with the observations. The zonal wind velocity being, by definition, constant with longitude, the only way to get around this is to introduce day/night asymmetries in the SSAS wind field. Precisely, near the equator, the SSAS wind needs to be stronger in the sub-solar region than in the anti-solar region. In this manner, the SSAS wind (resp. RSZ) dominates in the sub-solar region (resp. anti-solar region).\\
\\
We introduce such an asymmetry by modifying the SSAS wind-field on the day hemisphere. The proposed model is identical to the previous model with the exception that, on the day hemisphere, the SSAS wind velocity varies as V=V$_{ter}$*(1-$\big[\frac{\left|(90-sza)\right|}{90}\big]^5)$. This allows the SSAS wind velocity to drop less quickly away from the terminator, so that the SSAS wind is still significant even close to the eastern limb (sub-solar region). As a result, as shown in Figure \ref{combs} (bottom-right panel), the wind inversion is now detected only near the anti-solar region, as in the observations. However satisfying, this model can be still further adjusted so as to account for the remaining discrepancies to the data, that are mainly associated to observed asymmetries between Northern and Southern hemispheres. \\
\\
 To provide the best match to the data, we add to the previous model a latitudinal asymmetry in the SSAS wind-field, and use a parametrized expression to describe the SSAS wind velocity as V=V$_{ter}$*(1-$\big[\frac{\left|(90-sza)\right|}{90}\big]^X)$. At latitudes higher than 10$^{\circ}$ North, the best fitting SSAS wind model  appears to be very similar to the \citet{bougher1986} model, with only little day/night asymmetry. For June 12, the best results are obtained with X=1 on the day-side and X=0.8 on the night-side (resp. X=1.5 and X=1 on June 13). In contrast, for latitudes below 10$^{\circ}$ North, the best fitting SSAS wind model is flowing quickly on the day-side hemisphere, even close to the sub-solar region, while sharply slowing down past the terminator on the night-side hemisphere. That translates into a large variation of the X parameter between day and night hemispheres. The best results for June 12 are obtained for X=5 on the day-side and X=0.45 on the night-side (resp. X=5 and X=0.15 on June 13). The best fits for SSAS terminator velocity are obtained for V$_{ter}$=220~m/s (resp. V$_{ter}$=200~m/s), and, for the equatorial RSZ wind velocity, V$_{eq}$=100~m/s (resp. V$_{eq}$=70~m/s).  The best fitting wind models are graphically summarized on Figure \ref{proposedssas}, and the corresponding Doppler-shift maps are represented in Figure \ref{adjustedmaps}.\\
\\
These models provide a very nice match to the observed Doppler-shift distribution, but a global circulation modeling would be required to check if such spatial variations are theoretically possible and justified. In addition, these are certainly not the unique wind-field solution that could reproduce the observations. It is then difficult to quantify the temporal variations that can be derived from the observations. It is safe to say that,  in the context of the proposed models, the velocity of the  winds (both RSZ and SSAS) varies by as much as 30~m/s, although the wind-field structure does not change in the 24 hours separating the two observations.  \\

\section{Summary and discussion}

This paper presents observations of the CO(1-0) line on Venus obtained at the IRAM-Plateau de Bure interferometer in November 2007 and in June 2009. The measurements of the Doppler-shifts in the line core could map the upper mesosphere wind field on the morning hemisphere with a spatial resolution (3.5-5.5") much better than what can be reached with single-dish facilities. Atmospheric models based on the line-shape fitting allow one to estimate that the winds detected primarily sound altitudes between 96 and 101~km.\\ 
The Doppler-shift maps obtained reveal that the wind mostly follows a sky-East to West direction, corresponding in the observed geometry to the subsolar-to-antisolar (SSAS) wind direction. No significant meridional winds could be detected. In addition, an inversion of the flow direction is measured on the night-side near the equator. In 2009, the remarkable temporal stability between the two observing dates and the very good data quality gives strong confidence in these results.  The observations obtained in 2007 are of much lower quality mainly because of their poor imaging  quality  related to the near 0$^{\circ}$ declination of the target, and could not be analyzed in detail. The obtained wind pattern  however suggests some similarities with the 2009 data.\\ 
\\
To provide the best description of the wind field observed in 2009, we compared our results to a series of wind models. The best results are obtained with a global, dominant SSAS wind combined with an equatorial localized RSZ wind. Our proposed SSAS wind model includes significant asymmetries between the day and night hemispheres, as well as latitudinal variations. In particular, under 10$^{\circ}$ of latitude, the SSAS wind velocity decreases slowly across the disk from the terminator to the sub-solar region (day hemisphere), but decreases quickly from the terminator to the anti-solar region (night hemisphere). Recent observations by \citet{lellouch2008} and \citet{clancy2011} have also proposed latitudinal variations of the SSAS wind-field, but in a very different, latitudinally symmetric form, while our models point to significant North/South asymmetries.\\
Our SSAS wind model thus differs from the most commonly used model by \citet{bougher1986}, which proposes a linear variation of the wind velocity with solar incidence all over the disk. We notice that, although the wind model by \citet{bougher1986} is based on observational constraints of composition and temperature  distribution measured by Pioneer above 100~km, this model was never directly observationally confirmed. Recent airglow tracking \citep{hueso2008}  and mm-wave Doppler-shift mapping \citep{clancy2011} also revealed significant discrepancies between the observed wind field and the \citet{bougher1986} model. \\
The best fits obtained in our models lead to maximal SSAS wind velocities at the terminator of 200-220~m/s, which are comparable to the sound speed (200~m/s), but higher than most velocities estimated by previous works at this altitude, usually of the order of $\sim$100~m/s. However the same previous works did also display very large temporal variations in the SSAS wind velocity. Very high terminator velocities (290-322~m/s at 102~km) are compatible with the data presented in \citet{lellouch2008}, although they are not the favored solution. The SSAS wind modified model presented by \citet{clancy2011} is also characterized by low-latitude terminator velocities of $\sim$200~m/s at 108~km altitude. The velocities derived from this paper seem then plausible in the context of previous studies.  \\
Finally, our models propose the coexistence of a RSZ wind localized in an equatorial band (latitudes between -20$^{\circ}$ and 10$^{\circ}$), whose velocity at the equator varies between 70-100~m/s. The presence of a RSZ wind is in agreement with most observations of the upper mesosphere dynamics (mm-wave and IR Doppler-shift mapping, airglow and minor species distribution), where the SSAS wind is dominant and coexists with a very temporally variable RSZ wind.  However, the RSZ wind detected here is very spatially localized, which was never indicated by other observations. Only moderate variations of the zonal wind velocity with latitude were so far observed, for example in \citet{sornig2008,sornig2011}.  \\
Small temporal variations of the winds velocity by $\sim$30~m/s were also detected in both the SSAS wind and the RSZ wind in the 24~hour time span between the two 2009 observations, which is remarkably stable, since very large variations (up to 140~m/s for the RSZ wind and 60~m/s for the SSAS wind) could be observed on a similar timescale by \citet{clancy2011},  and even more radical changes in the wind structure can be monitored on longer timescales.\\
\\
Although not unambiguous, our interpretation of the wind field observed in 2009 is in agreement with the usual picture of the upper mesosphere dynamics (combination of SSAS and RSZ winds), but at the same time clearly indicates that the wind field structure is more complicated than the usually adopted models. This work demonstrates the ability of mm-wave interferometry to characterize a wind field, on a larger spatial scale and with an easier interpretation than the airglow tracking method. In order to describe the wind field on a planet-wide scale, and to constrain the long term stability of the wind structures, observations at other venusian phases are needed,  especially near superior conjunction,
when the entire day-side disk is sampled and Venus' apparent size is minimal ($\sim$10"). Future studies accessible with the Atacama Large Millimeter Array (ALMA) could include high spatial resolution mapping ($\sim$1") aimed at exploring dynamics in the high latitude regions and short timescale ($\sim$1~hour) monitoring to detect quick changes in the wind structure and velocity.

\begin{acknowledgements}
We thank T. Clancy for providing us with the reference atmospheric profiles.
We acknowledge the commitment of the staff at IRAM-PdBI to perform these challenging observations.
IRAM is supported by INSU/CNRS (France), MPG (Germany) and IGN (Spain).
\end{acknowledgements}

\bibliographystyle{aa} 

\begin{table*}
\begin{tabular}{|c|c|c|c|c|c|c|}
\hline
Date & Equatorial Diameter & Local Time at Central & Observation Length & T$_{sys}$ at Transit & Beam Size & Beam Position \\
 & (") &Meridian (hours) &  (hours) & (K) & (") & Angle ($^{\circ}$)\\
\hline
16 and 17 November, 2007 & 20.0 & 6.8 am & 16 & 350 & 5.7$\times$3.6 & -17\\
12 June, 2009 & 22.1 & 6.2 am & 6.5 & 300 & 4.1$\times$4.1 & -27\\
13 June, 2009 & 21.9 & 6.2 am & 8 & 300 & 5.7$\times$5.3 & -44\\
\hline
\end{tabular}
\caption{\label{ephem} Observational parameters of Venus at the observation dates. From www.imcce.fr.}
\end{table*}

\end{document}
\fi